\documentclass{article}

\usepackage{arxiv}

\usepackage[utf8]{inputenc} 
\usepackage[T1]{fontenc}    
\usepackage{hyperref}       
\usepackage{url}            

\usepackage{amsfonts}       
\usepackage{nicefrac}       
\usepackage{microtype}      
\usepackage{lipsum}

\usepackage[numbers]{natbib}
\usepackage{booktabs}
\usepackage{multirow}
\usepackage{algorithm}
\usepackage{algorithmic}
\usepackage{graphicx} 
\usepackage{subcaption} 
\newtheorem{theorem}{Theorem}
\newtheorem{definition}{Definition}
\usepackage{amsmath}
\usepackage{xcolor}
\usepackage{soul}

\makeatletter
\renewcommand{\fnum@table}{\fontfamily{ptm}\selectfont Table~\thetable}
\AtBeginDocument{%
	\let\oldcaption\caption
	\renewcommand{\caption}[2][]{\oldcaption[#1]{\fontfamily{ptm}\selectfont #2}}
}
\makeatother

\title {ICBAC: an Intelligent Contract-Based Access Control framework for supply chain management by integrating blockchain and federated learning}

\author{
	Sadegh Sohani \\
	Department of Computer Engineering \\
	Ferdowsi University of Mashhad \\
	Mashhad, Iran \\
	\texttt{sd\_sohani@mail.um.ac.ir} \\
	\And
	Salar Ghazi \\
	École de technologie supérieure \\
	Montréal, Québec, Canada \\
	\texttt{seyed-salar.ghazi.1@ens.etsmtl.ca} \\
	\And
	Farnaz Kamranfar \\
	Department of Computer Engineering \\
	Ferdowsi University of Mashhad \\
	Mashhad, Iran \\
	\texttt{FarnazKamranfar1999@gmail.com} \\
	\And
	Sahar Pilehvar Moakhar \\
	Department of Computer Engineering \\
	Ferdowsi University of Mashhad \\
	Mashhad, Iran \\
	\texttt{sahar.pilevarmoakhar@mail.um.ac.ir} \\
	\And
	Mohammad Allahbakhsh \\
	Department of Computer Engineering \\
	Ferdowsi University of Mashhad \\
	Mashhad, Iran \\
	\texttt{allahbakhsh@um.ac.ir} \\
	\And
	Haleh Amintoosi \\
	Department of Computer Engineering \\
	Ferdowsi University of Mashhad \\
	Mashhad, Iran \\
	\texttt{amintoosi@um.ac.ir} \\
	\And
	Kaiwen Zhang \\
	École de technologie supérieure \\
	Montréal, Québec, Canada \\
	\texttt{kaiwen.zhang@etsmtl.ca} \\
}

\begin{document}
\maketitle
\begin{abstract}
Modern supply chain management (SCM) systems operate across multiple independent and often competing organizations, making access control a critical yet unresolved challenge. Existing access control mechanisms in SCM are predominantly static and centralized, limiting their ability to adapt to insider threats, erroneous behaviors, and evolving operational contexts. Although blockchain technology improves transparency and decentralization, most blockchain-based SCM access control frameworks still rely on predefined policies and lack behavioral intelligence. At the same time, centralized machine learning solutions for anomaly detection require aggregating sensitive behavioral data, which is incompatible with the privacy and competitiveness constraints of real-world supply chains.

This paper proposes \textit{ICBAC}, an intelligent, contract-based access control framework for multi-party supply chains that integrates permissioned blockchain technology with federated learning (FL). Built on Hyperledger Fabric, ICBAC enforces access policies through a multi-channel architecture and three dedicated smart contracts for asset management, baseline access control, and dynamic permission revocation. To address insider misuse and anomalous behavior, each supply chain channel deploys an AI agent that monitors participant activity and dynamically restricts access when anomalies are detected. Federated learning enables these agents to collaboratively improve anomaly detection models without sharing raw data, preserving confidentiality across competing supply chains.

To account for heterogeneous and competitive environments, ICBAC introduces a game-theoretic client selection mechanism based on hedonic coalition formation. This mechanism allows supply chains to form stable and strategy-proof federated learning coalitions using preference-based selection, without disclosing sensitive collaboration criteria. Extensive experiments conducted on a Hyperledger Fabric testbed and a real-world supply chain dataset demonstrate that ICBAC achieves blockchain performance comparable to state-of-the-art static frameworks, while providing effective anomaly detection under both IID and non-IID data distributions with zero raw-data sharing. The results indicate that ICBAC offers a practical and scalable solution for dynamic, privacy-preserving access control in decentralized supply chain ecosystems.
\end{abstract}


\section{Introduction}

Supply Chain Management (SCM) has evolved into a mission-critical function for enterprises worldwide, yet it remains vulnerable to security and operational risks. The global SCM software market is projected to grow from \$29 billion in 2023 to \$62 billion by 2028, driven by large-scale modernization initiatives and investments in advanced supply chain technologies \cite{Gartner2024}. Rising renewal pricing for market leaders further reflects the increasing strategic importance of SCM platforms. This rapid growth amplifies both the value of supply chain data and the consequences of failures, underscoring the urgency of addressing inherent security and governance challenges.

Despite its strategic role, SCM remains heavily dependent on human-driven and cross-organizational processes, exposing organizations to significant operational and security risks. Inadequate procedures, IT outages, human errors, staffing shortages, and infrastructure failures frequently disrupt supply chain operations, often exacerbated by insufficient or poorly enforced access control mechanisms \cite{DakhchBenamar2024,WagnerBode2006}. In parallel, internal cyberattacks—where authorized insiders abuse legitimate privileges—have emerged as one of the most costly threat vectors in recent years, with average breach costs reaching several million USD. Supply chain breaches involving third-party vendors have also increased in prevalence and impact, accounting for a substantial fraction of incidents and requiring extended periods for detection and containment \cite{Kshetri2022,IBM2024,IBM2025}. Unlike external attacks that exploit software vulnerabilities, these incidents frequently leverage legitimate access rights, making them particularly difficult to detect using static authorization policies.

These challenges highlight the need for access control mechanisms that are not only decentralized and auditable, but also adaptive to behavioral misuse in multi-party environments. Traditional centralized systems that aggregate sensitive organizational data struggle to balance transparency, trust, and confidentiality across independent participants. Blockchain technology offers a promising alternative by providing a distributed and immutable ledger secured through cryptographic hashing and consensus protocols \cite{Nakamoto2008,Sohani2024,tbac}. By enabling smart contracts to automate and transparently enforce policies without centralized intermediaries, blockchain fosters trust and accountability among mutually distrustful organizations \cite{shahcons,SmartContract2025}. As a result, blockchain-based access control has attracted increasing attention in SCM contexts.

However, blockchain-based enforcement alone is insufficient to address insider threats and erroneous behaviors. Smart contracts can reliably enforce predefined rules, but they lack the ability to interpret evolving behavioral patterns. Access control defines which entities are authorized to view or manipulate system resources \cite{AccessControl2025}, yet static policies cannot capture context-dependent misuse. Machine learning–based User and Entity Behavior Analytics (UEBA) provides a complementary capability by detecting anomalies through behavioral analysis in real time \cite{PaloAlto2025}. In supply chain environments, behavioral data is particularly informative because users with legitimate access privileges may still exhibit anomalous patterns—such as unusual timing, frequency, or magnitude of operations—that indicate malicious abuse or unintentional errors. For example, in warehouse management systems, multiple agents may be authorized to modify inventory levels, yet certain behavioral signatures signal misuse that static rules cannot capture.

Training AI models to recognize such anomalies requires learning from historical behavioral data, as malicious patterns such as credential abuse, unusual access timing, or bulk operations often exhibit recurring signatures. However, sharing raw behavioral data across supply chains and organizations introduces significant privacy, regulatory, and competitive risks. Federated learning (FL) addresses this challenge by enabling multiple parties to collaboratively train models without exchanging underlying data, thereby preserving confidentiality while leveraging distributed behavioral insights \cite{Federated-Learning}. Nevertheless, applying FL directly at large scale across heterogeneous and competing supply chains introduces new challenges related to data heterogeneity, convergence instability, and strategic participation.

Recent studies have explored integrating blockchain and federated learning into access control frameworks for domains such as the Internet of Things (IoT) and telecommunications \cite{Cao2022,Lin2023,Wang2024}. While these approaches demonstrate the feasibility of decentralized and privacy-aware learning, most assume cooperative settings or uniform objectives. Such assumptions do not hold in SCM, where organizations may be competitors and where behavioral norms vary significantly across supply chains. In particular, behaviors considered anomalous in one supply chain may be entirely normal in another, making global or fully centralized federated learning both ineffective and potentially harmful.

To address these challenges, this paper proposes the \emph{Intelligent Contract-Based Access Control} (ICBAC) framework for secure and privacy-preserving supply chain management. ICBAC is built on Hyperledger Fabric, a permissioned blockchain platform that supports private channels for selective data sharing among authorized organizations. The use of multiple channels reflects a key design choice: each channel corresponds to a distinct supply chain context with its own operational semantics, trust boundaries, and behavioral norms. This separation enables fine-grained governance and prevents behavioral interference across unrelated supply chains.

Within each channel, ICBAC deploys a single AI agent responsible for monitoring participant behavior and detecting anomalies. This design choice is motivated by both scalability and accountability considerations. A per-channel AI agent allows anomaly detection models to specialize to the behavioral characteristics of a specific supply chain, avoiding the dilution effects of global models trained on heterogeneous data. At the same time, it limits computational and coordination overhead, enabling practical deployment in real-world systems where resources and administrative capacity are constrained.

To improve anomaly detection without violating privacy constraints, AI agents across channels participate in federated learning. Rather than forming a single large-scale federation, ICBAC adopts a coalition-based FL approach. Because data distributions, operational practices, and threat models differ across supply chains, learning across all participants can degrade model performance and convergence. ICBAC therefore introduces a client-selection mechanism grounded in hedonic coalition-formation game theory, allowing channels with compatible behavioral patterns to form stable, small-scale federated learning coalitions. This mechanism is executed once—after a sufficient number of candidates is available and before initiating federated learning—to establish stable collaboration groups for the entire training process.

Coalitions are formed based on preference declarations expressed as lists of trusted or compatible partners, without revealing the sensitive criteria underlying these preferences. By leveraging Tarjan’s algorithm to identify strongly connected components in the resulting preference graph, the mechanism provides formal guarantees of core stability and strategy-proofness, ensuring that no group of agents has an incentive to deviate and that agents cannot benefit from misrepresenting their preferences. This design preserves privacy, limits strategic manipulation, and enables efficient federated learning among behaviorally compatible supply chains.

The main contributions of this paper are summarized as follows:
\begin{itemize}
	\item \textbf{Blockchain-Based Access Control Framework for SCM:} We propose ICBAC, a secure and privacy-preserving access control framework for supply chain management built on Hyperledger Fabric, employing a multi-channel architecture to reflect heterogeneous supply chain contexts.
	\item \textbf{AI-Driven Dynamic Access Revocation:} We integrate per-channel AI agents to monitor participant behavior and detect anomalies, enabling automated and context-aware permission revocation through smart contracts.
	\item \textbf{Privacy-Preserving Federated Learning:} We employ federated learning to allow AI agents to collaboratively improve anomaly detection models without sharing sensitive behavioral data.
	\item \textbf{Game-Theoretic Client Selection Mechanism:} We introduce a hedonic coalition-formation–based client selection mechanism that enables stable, small-scale federated learning while ensuring core stability, strategy-proofness, and privacy preservation.
\end{itemize}

The remainder of this paper is structured as follows. Section~\ref{sec:lr} reviews related work and provides a comparative analysis. Section~\ref{sec:prom} presents the proposed ICBAC framework and its components. Section~\ref{sec:eval} evaluates the system through extensive experiments, and Section~\ref{sec:con} concludes the paper with key findings and future research directions.

\section{Related Work}
\label{sec:lr}

This section reviews existing access control models and frameworks relevant to supply chain management (SCM). Section~\ref{ssec:TACM} discusses traditional centralized access control approaches adapted to SCM. Section~\ref{ssec:BBAC} reviews decentralized and blockchain-based access control models proposed for supply chain environments. Section~\ref{ssec:AD-SCM} examines blockchain and federated learning (FL)–based access control frameworks from other domains to highlight their relevance and limitations when applied to SCM.

\subsection{Traditional Access Control Models}
\label{ssec:TACM}

Early access control solutions for SCM primarily adapted traditional models such as Role-Based Access Control (RBAC) \cite{Ravi1998} and Attribute-Based Access Control (ABAC) \cite{Hu2015}. These approaches focus on defining static authorization policies based on user roles or attributes and have been widely adopted due to their conceptual simplicity and compatibility with enterprise systems.

Several studies extended these models to address specific SCM challenges. Gao et al.~\cite{Gao2004} proposed a randomized access control mechanism to enhance RFID security and privacy, while Zhang and Li~\cite{Zhang2006} examined secure information sharing using RBAC and collaborative protocols in Internet-based SCM. Auditing-based mechanisms were explored in~\cite{Bharat2013} to improve accountability and detect unauthorized disclosure in distributed supply chains. RFID-centric access control and authentication schemes were further investigated in~\cite{Yang2015,Qi2016}, enabling fine-grained access control and scalable privilege management. More recent work integrates biometric and multi-factor authentication to enhance usability and security~\cite{Lee2021}.

Despite these efforts, traditional access control models remain fundamentally centralized and policy-driven. They assume static trust relationships and lack mechanisms to adapt to evolving behavioral patterns or insider misuse. As a result, these approaches struggle to address dynamic, multi-party SCM environments where authorized users may still pose security risks.

\subsection{Blockchain-Based Access Control Models}
\label{ssec:BBAC}

To overcome the limitations of centralized access control, numerous studies have explored blockchain-based solutions for SCM. Blockchain’s decentralized architecture, immutability, and smart contract support enable transparent and tamper-resistant policy enforcement across organizational boundaries.

Several frameworks leverage blockchain to enhance identity management and access control. BIMAC~\cite{Liao2022} and multi-authority attribute-based schemes~\cite{Liu2022} enable privacy-preserving data sharing in decentralized environments. Hybrid RBAC–ABAC models integrated with Hyperledger Fabric have been proposed for medical and pharmaceutical supply chains~\cite{Li2023,Hathaliya2024,Sharma2024}. Ethereum-based systems such as AccessChain~\cite{Sarfaraz2023} introduce decentralized access policy management using attribute-based control.

Other studies focus on improving privacy and scalability through cryptographic techniques and off-chain storage. ProChain~\cite{Li2024} employs zero-knowledge proofs and CP-ABE for privacy-preserving traceability, while IPFS-based architectures enhance data availability and storage efficiency~\cite{Raj2024,Sohani2024,Li2025}. Blockchain-based access control has also been applied to agricultural, food, port, and vaccine supply chains to improve traceability and operational transparency~\cite{Dash2024,Rahaman2024,Jha2025,Arsheen2025}. Dynamic access control mechanisms have been explored in collaborative emergency management~\cite{Wang2025}.

Although blockchain-based access control frameworks improve decentralization, transparency, and auditability, most rely on predefined authorization policies and static rule enforcement. They rarely incorporate behavioral intelligence or adaptive mechanisms to mitigate insider misuse, erroneous operations, or context-dependent threats. Consequently, these approaches remain limited in addressing the dynamic and behavior-driven security challenges of modern SCM.

\subsection{Blockchain and Federated Learning–Based Access Control in Other Domains}
\label{ssec:AD-SCM}

The integration of machine learning with blockchain-based access control has gained traction in domains such as IoT, healthcare, and networked systems. Federated learning (FL) is particularly well aligned with blockchain due to its decentralized and privacy-preserving training paradigm \cite{Federated-Learning,FL-blockchain}. While this integration has not been explicitly explored for SCM access control, related work in other domains provides valuable insights.

Several studies employ FL to enhance access control and resource management. In telecommunications, federated deep reinforcement learning has been applied to optimize access decisions in Open RAN systems~\cite{Cao2022}. Blockchain-assisted FL frameworks have been proposed to prevent model poisoning and ensure trustworthy aggregation in industrial IoT~\cite{Kalapaaking2023,Singh2023}. Privacy-aware access control mechanisms combining FL with deep learning and trust evaluation have been explored in IoT-based healthcare~\cite{Lin2023}.

More recent work investigates scalable FL architectures and security enhancements. Resource-optimized FL frameworks using reinforcement learning~\cite{Mishra2024}, hierarchical blockchain-enabled FL~\cite{Wang2024}, and blockchain-integrated FL systems for IoMT~\cite{Ramani2024,Jafari2025} address challenges such as non-IID data, privacy leakage, and system scalability. Additional studies incorporate explainable AI, edge computing, and advanced communication middleware to improve interpretability, efficiency, and robustness~\cite{Ali2025,Gu2025,Ababio2025,Teixeira2025,Hoang2025}.

While these approaches demonstrate the feasibility of combining blockchain and FL for intelligent and privacy-aware systems, they largely assume cooperative environments with aligned objectives. They do not address the competitive, multi-organizational nature of SCM, where participants may have conflicting interests and heterogeneous behavioral norms. As a result, existing FL–blockchain frameworks are not directly applicable to SCM access control without significant adaptation.

\subsection{Discussion}

Overall, traditional access control models adapted to SCM suffer from centralization, limited transparency, and an inability to adapt to evolving behaviors~\cite{Gao2004,Zhang2006,Bharat2013}. Blockchain-based frameworks improve decentralization and auditability but typically enforce static policies without behavioral intelligence~\cite{Liao2022,Liu2022,Sarfaraz2023}. Although FL–blockchain integration has been explored in healthcare, IoT, and industrial systems~\cite{Singh2023,Hoang2025}, these solutions do not account for the competitive dynamics, heterogeneous behaviors, and privacy constraints inherent to SCM.

ICBAC addresses these gaps by combining blockchain-based access control with AI-driven behavioral anomaly detection and federated learning specifically tailored to supply chain environments. Unlike existing frameworks, ICBAC supports dynamic permission adaptation, promotion and demotion of privileges, and privacy-preserving collaboration through a game-theoretic coalition formation mechanism. This design enables intelligent, decentralized, and context-aware access control suitable for modern, multi-party supply chain ecosystems.

\section{Framework Model Design} \label{sec:prom}

\begin{table}[!h]
	\fontfamily{ptm}\selectfont
	\caption{Table of notations.}
	\centering
	\label{tab:notations}
	\begin{tabular}{ll}
		\toprule
		\textbf{Symbol} & \textbf{Description} \\
		\midrule
		$P_j$ & Participant $j$ in the supply chain \\
		$A_j$ & Asset $j$ being accessed/modified \\
		$o_k$ & Operation $k$ (e.g., read, write, update) \\
		$v$ & Value/data for the operation \\
		$C_i$ & Channel $i$ representing a supply chain \\
		$PRL_i$ & Permission Revoke List for channel $i$ \\
		$\mathcal{L}_{j,j}$ & Least privilege set for participant $P_j$ on asset $A_j$ \\
		$\mathcal{E}_{j,j}$ & Extra privilege set for participant $P_j$ on asset $A_j$ \\
		$\mathcal{Q}_{j,j,k}$ & Allowed operations on attribute $attr_k$ under least privilege \\
		$\mathcal{R}_{j,j,k}$ & Allowed operations on attribute $attr_k$ under extra privilege \\
		$\mathbf{x}_t$ & Feature vector of an access event at time $t$ \\
		$\mathbf{X}_t$ & Windowed sequence of events of length $T$ ending at time $t$ \\
		$T$ & Sequence/window length for temporal analysis \\
		$AI_i$ & AI agent monitoring channel $C_i$ \\
		$\theta_i$ & Local model parameters for channel $C_i$ \\
		$\theta$ & Aggregated model parameters \\
		$\mathcal{D}_i$ & Local dataset for channel $C_i$ \\
		$F(\succeq_{AI_i})$ & Friendship list (preferences) of $AI_i$ \\
		$\Gamma$ & Friendship graph \\
		$\mathcal{CS}$ & Coalition of channels participating in federated learning \\
		$N_i$ & Number of samples in channel $C_i$ \\
		$w_i$ & FedAvg weight for channel $C_i$ \\
		\bottomrule
	\end{tabular}
\end{table}

This section presents the architecture and core components of ICBAC. Section~\ref{ss:sa} introduces the system architecture. Section~\ref{ss:cid} details channels, assets, smart contracts, AI agents, federated learning, and client selection. Section~\ref{ss:ow} summarizes the end-to-end workflow. Table~\ref{tab:notations} lists the notations used in this section.

\subsection{Architecture}\label{ss:sa}

\begin{figure*}[!t]
	\centering
	\includegraphics[scale=0.5]{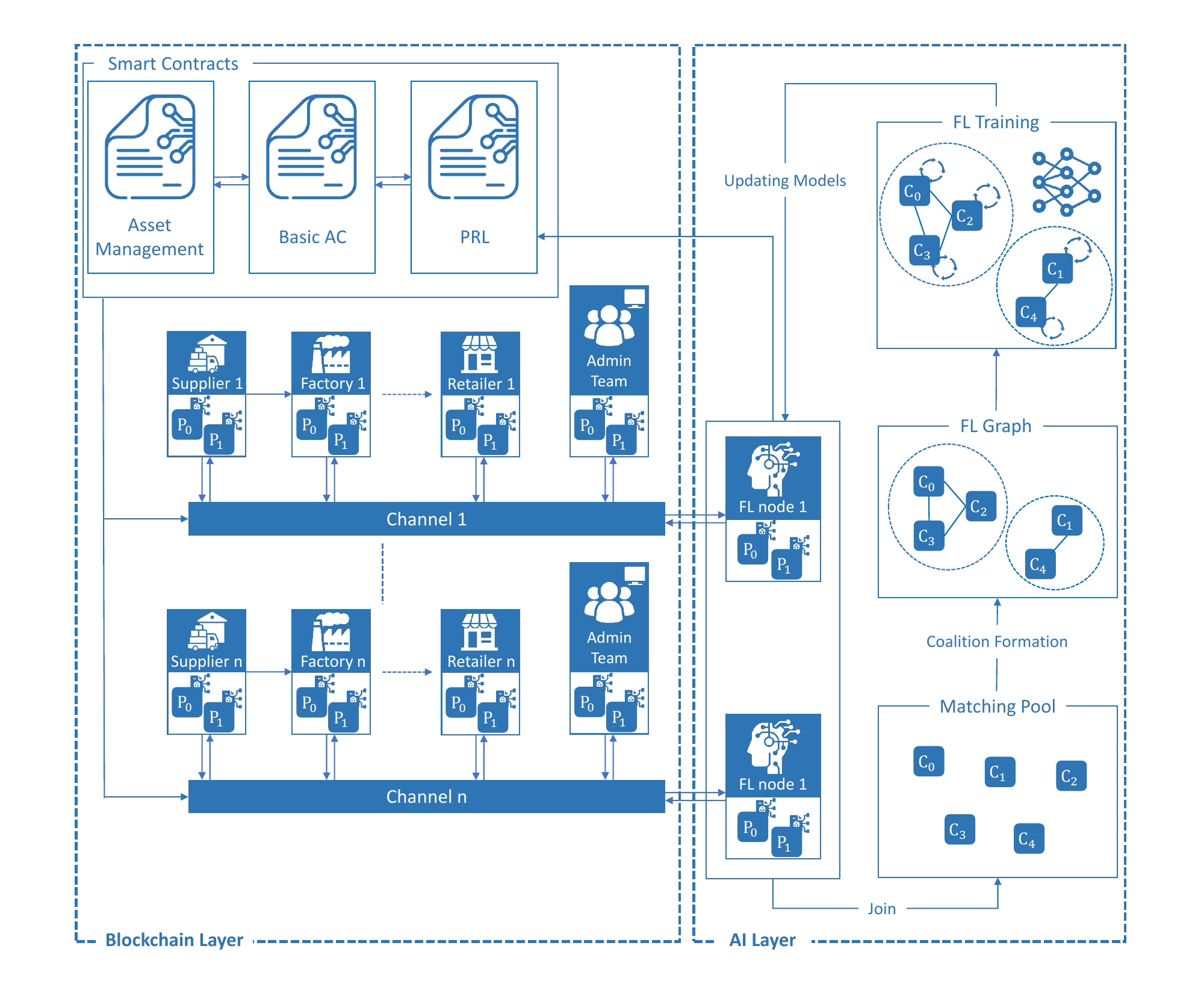}
	\caption{Overall architecture of the proposed ICBAC framework.}
	\label{fig:arch}
\end{figure*}

As shown in Figure~\ref{fig:arch}, ICBAC consists of two layers: (i) a blockchain layer and (ii) an artificial intelligence layer. The blockchain layer provides authenticated membership, tamper-evident logging, and smart contract execution for access control and asset operations. We adopt Hyperledger Fabric \cite{Hyperledger} because it is a permissioned blockchain that supports private channels for selective data sharing among authorized organizations, which is aligned with confidentiality requirements in SCM. Fabric’s modular design and chaincode execution model also facilitate implementing fine-grained access logic and auditability.

The AI layer provides adaptive security by monitoring participant behaviors and supporting anomaly-driven access restriction. Each Fabric channel is associated with one AI agent that learns channel-specific behavioral norms. When suspicious behavior is detected, the AI agent proposes a temporary access restriction by inserting the participant into the Permission Revoke List (PRL), subject to administrative review. To improve anomaly detection without cross-channel raw data sharing, AI agents participate in federated learning (FL). Since behavioral norms differ across supply chains, ICBAC incorporates a client selection mechanism that forms coalitions of behaviorally compatible channels for FL.

We formalize ICBAC as:
\begin{flalign}
	\mathcal{S} = (\mathcal{C}, \mathcal{P}, \mathcal{A}, \mathcal{SC}, \mathcal{AI}, \mathcal{FL}, \mathcal{CS})
\end{flalign}
where:
\begin{itemize}
	\item $\mathcal{C} = \{C_1, \ldots, C_m\}$ is the set of channels (each representing a supply chain context).
	\item $\mathcal{P} = \{P_1, \ldots, P_n\}$ is the set of participants.
	\item $\mathcal{A} = \{A_1, \ldots, A_k\}$ is the set of managed assets.
	\item $\mathcal{SC} = \{\mathcal{SC}_{AM}, \mathcal{SC}_{BAC}, \mathcal{SC}_{PRL}\}$ is the set of smart contracts for asset management, access control, and revocation.
	\item $\mathcal{AI} = \{AI_1, \ldots, AI_m\}$ is the set of per-channel AI agents.
	\item $\mathcal{FL}$ denotes the federated learning procedure.
	\item $\mathcal{CS}$ denotes the client selection mechanism that determines coalitions for FL.
\end{itemize}

\subsection{ICBAC Components in Detail} \label{ss:cid}

\subsubsection{Channel and Participant Structure}

Each channel $C_i \in \mathcal{C}$ represents a distinct supply chain context with its own participants and assets, enabling isolation and selective disclosure across contexts. Let $\mathcal{P}_i \subseteq \mathcal{P}$ and $\mathcal{A}_i \subseteq \mathcal{A}$ denote the participant and asset sets of channel $C_i$:
\begin{flalign}
	\mathcal{P}_i = \{P_j \in \mathcal{P} \mid P_j \text{ is an authorized member of } C_i\}
\end{flalign}
\begin{flalign}
	\mathcal{A}_i = \{A_k \in \mathcal{A} \mid A_k \text{ is managed within } C_i\}.
\end{flalign}

\subsubsection{Asset Management Smart Contract (\(\mathcal{SC}_{AM}\))}

The Asset Management smart contract governs asset creation and updates. Each asset $A_j \in \mathcal{A}$ is represented by attributes:
\begin{flalign}
	A_j = \{attr_1, attr_2, \ldots, attr_{n_j}\}.
\end{flalign}
Let $\mathcal{O}=\{o_1,\ldots,o_l\}$ denote allowable operations (e.g., read, write, update). A request is a tuple $(P_j, A_j, o_k, v)$ and is evaluated against access policies enforced by $\mathcal{SC}_{BAC}$ and eligibility checks enforced by $\mathcal{SC}_{PRL}$.

\subsubsection{Basic Access Control Smart Contract (\(\mathcal{SC}_{BAC}\))}

$\mathcal{SC}_{BAC}$ manages baseline and dynamically adjusted permissions through two collections: \textit{LeastPrivilege} and \textit{ExtraPrivilege}:

\begin{align}
	\textit{LeastPrivilege} &= \{ (P_j, A_j) \mapsto \mathcal{L}_{j,j} \}, \notag \\
	\mathcal{L}_{j,j} &= \{ (attr_k, \mathcal{Q}_{j,j,k}) \mid attr_k \in A_j,\; \mathcal{Q}_{j,j,k} \subseteq \mathcal{O} \}
\end{align}

\begin{align}
	\textit{ExtraPrivilege} &= \{ (P_j, A_j) \mapsto \mathcal{E}_{j,j} \}, \notag \\
	\mathcal{E}_{j,j} &= \{ (attr_k, \mathcal{R}_{j,j,k}) \mid attr_k \in A_j,\; \mathcal{R}_{j,j,k} \subseteq \mathcal{O} \}.
\end{align}

To index permission records, we compute:
\begin{flalign}\label{eq:sha}
	\textit{Index}(P_j, A_j) = \textit{SHA256}(P_j \parallel A_j).
\end{flalign}

An operation $o_k$ on attribute $attr_k$ of $A_j$ requested by participant $P_j$ is allowed by $\mathcal{SC}_{BAC}$ if:
\begin{flalign}
	o_k \in \mathcal{Q}_{j,j,k} \cup \mathcal{R}_{j,j,k}.
\end{flalign}

Privilege updates are performed via promotion/demotion on $\mathcal{R}_{j,j,k}$ (extra privileges):
\begin{flalign}
	\textit{Promote}: (P_j, A_j, attr_k, o_k) \to \mathcal{R}_{j,j,k} \cup \{o_k\}
\end{flalign}
\begin{flalign}
	\textit{Demote}: (P_j, A_j, attr_k, o_k) \to \mathcal{R}_{j,j,k} \setminus \{o_k\}.
\end{flalign}

\subsubsection{Permission Revoke List Smart Contract (\(\mathcal{SC}_{PRL}\))}

$\mathcal{SC}_{PRL}$ supports anomaly-driven temporary restriction by maintaining a per-channel revoke list $\mathcal{PRL}_i$ and a pending approval queue:
\begin{flalign}
	\mathcal{PRL}_i &= \{P_j \in \mathcal{P}_i \mid P_j \text{ is flagged as anomalous by } AI_i\}
\end{flalign}

\begin{flalign}
	\textit{PendingRequests}_i
	&= \{(P_j, A_j, attr_k, o_k, v) \mid \notag \\
	&\quad \text{the request awaits approval by } \mathcal{T}_i \}. \label{eq:pending}
\end{flalign}

where $\mathcal{T}_i \subseteq \mathcal{P}_i$ is the admin team of channel $C_i$.

Participants are inserted/removed as:
\begin{flalign}
	\textit{AddPRL}: P_j \to \mathcal{PRL}_i \cup \{P_j\},
\end{flalign}
\begin{flalign}
	\textit{RemovePRL}: P_j \to \mathcal{PRL}_i \setminus \{P_j\}.
\end{flalign}

An access request is processed as:
\begin{equation}
	\label{eq:access}
	\textit{Access}(P_j, A_j, attr_k, o_k)=
	\begin{cases}
		\text{Pending}, & \text{if } P_j \in \mathcal{PRL}_i,\\[2pt]
		\text{Deny}, & \begin{aligned}[t]
			\text{if } & P_j \notin \mathcal{PRL}_i \\
			& \land\ o_k \notin \mathcal{Q}_{j,j,k} \cup \mathcal{R}_{j,j,k},
		\end{aligned}\\[2pt]
		\text{Allow}, & \begin{aligned}[t]
			\text{if } & P_j \notin \mathcal{PRL}_i \\
			& \land\ o_k \in \mathcal{Q}_{j,j,k} \cup \mathcal{R}_{j,j,k}.
		\end{aligned}
	\end{cases}
\end{equation}

Algorithm~\ref{alg:basic-ac} summarizes the access decision logic.

\begin{algorithm}
	\caption{Check Access Procedure (consistent with $\mathcal{SC}_{BAC}$ and $\mathcal{SC}_{PRL}$)}
	\label{alg:basic-ac}
	\begin{algorithmic}[1]
		\REQUIRE Participant $P_j$, Asset $A_j$, Attribute $attr_k$, Operation $o_k$, Channel $C_i$
		\ENSURE Access decision: Allow / Deny / Pending
		\STATE Let $\mathcal{P}_i$ be the participant set of $C_i$
		\STATE Let $\mathcal{PRL}_i$ be the revoke list of $C_i$
		\STATE Let $\mathcal{Q}_{j,j,k}$ and $\mathcal{R}_{j,j,k}$ be the least and extra allowed operation sets for $(P_j,A_j,attr_k)$ in $\mathcal{SC}_{BAC}$
		
		\IF{$P_j \notin \mathcal{P}_i$}
		\RETURN Deny \COMMENT{Not a member of the channel}
		\ENDIF
		
		\IF{$P_j \in \mathcal{PRL}_i$}
		\RETURN Pending \COMMENT{Temporarily restricted; admin review required}
		\ENDIF
		
		\IF{$o_k \notin \mathcal{Q}_{j,j,k} \cup \mathcal{R}_{j,j,k}$}
		\RETURN Deny \COMMENT{Not authorized under least or extra privileges}
		\ENDIF
		
		\RETURN Allow
	\end{algorithmic}
\end{algorithm}

\subsubsection{AI-Based Behavioral Anomaly Detection}
\label{ss:ai_anomaly}

The AI component of ICBAC is responsible for detecting anomalous access behaviors that may indicate insider misuse, erroneous operations, or policy violations not captured by static access rules. To ensure scalability and deployability across heterogeneous supply chain environments, anomaly detection is performed locally within each blockchain channel using a lightweight temporal model trained on channel-specific behavioral data.

\paragraph{Behavioral Representation.}
Access activities are represented as time-ordered sequences of events generated by participant interactions with assets. Each access event at time $t$ is encoded as a feature vector
\begin{equation}
	\mathbf{x}_t \in \mathbb{R}^d,
\end{equation}
where features may include participant identity, asset type, operation type, access frequency, temporal context, and other behavioral attributes. To capture temporal dependencies, a sliding window of length $T$ is used to construct behavioral sequences:
\begin{equation}
	\mathbf{X}_t = \{\mathbf{x}_{t-T+1}, \ldots, \mathbf{x}_t\}.
\end{equation}

\paragraph{Model Abstraction.}
Each channel deploys a temporal anomaly detection model that learns normal behavioral patterns from historical access sequences. The model maps an input sequence $\mathbf{X}_t$ to an expected representation of normal behavior:
\begin{equation}
	\hat{\mathbf{x}}_t = f(\mathbf{X}_t; \boldsymbol{\theta}),
\end{equation}
where $f(\cdot)$ denotes a channel-specific temporal prediction function parameterized by $\boldsymbol{\theta}$. The proposed framework is intentionally agnostic to the specific realization of $f(\cdot)$, allowing different lightweight sequence models to be adopted depending on operational constraints, computational resources, and deployment requirements.

\paragraph{Training Objective.}
The model is trained to minimize prediction error on predominantly normal behavior, enabling deviations from learned patterns to be identified as anomalies. The loss function for an access event at time $t$ is defined as:
\begin{equation}
	\mathcal{L}_t = \left\| \mathbf{x}_t - \hat{\mathbf{x}}_t \right\|_2^2.
\end{equation}
Training minimizes the empirical risk over local channel data:
\begin{equation}
	\min_{\boldsymbol{\theta}} \; \frac{1}{N} \sum_{t=1}^{N} \mathcal{L}_t,
\end{equation}
where $N$ is the number of training samples in the channel.

\paragraph{Federated Training.}
To improve anomaly detection performance while preserving data confidentiality, model parameters are optionally refined using federated learning across a coalition of channels $\mathcal{CS} \subseteq \mathcal{C}$. Each channel $C_i$ performs local training on its dataset $\mathcal{D}_i$ and shares model updates rather than raw data. Aggregation at round $k{+}1$ follows:
\begin{equation}
	\boldsymbol{\theta}^{(k+1)} = \sum_{i \in \mathcal{CS}} w_i \boldsymbol{\theta}_i^{(k)},
\end{equation}
where $w_i$ is proportional to the local sample size $N_i$ of channel $C_i$ \cite{Federated-Learning}.

\paragraph{Inference and Anomaly Scoring.}
During inference, the anomaly score for an access event at time $t$ is computed as:
\begin{equation}
	s_t = \left\| \mathbf{x}_t - \hat{\mathbf{x}}_t \right\|_2^2.
\end{equation}
An event is flagged as anomalous if $s_t$ exceeds a predefined threshold $\tau$, selected empirically based on validation data:
\begin{equation}
	s_t > \tau.
\end{equation}
When anomalies are detected, the corresponding participant is temporarily added to the Permission Revoke List $\mathcal{PRL}_i$ of the channel, and subsequent access requests are marked as \emph{Pending} until reviewed by the channel administrator team.

\subsubsection{Federated Learning (\(\mathcal{FL}\))}

In ICBAC, federated learning enables AI agents across channels to improve anomaly detection without sharing raw behavioral or transactional data across supply chains. Each channel $C_i \in \mathcal{C}$ has an AI agent $AI_i$ with local parameters $\theta_i$ trained on its channel-specific sequence data (Section~\ref{ss:ai_anomaly}). We adopt Federated Averaging (FedAvg) \cite{FedAVG}. A subset of channels $\mathcal{CS} \subseteq \mathcal{C}$ is selected by the client selection mechanism, and each agent performs local updates before aggregation:
\begin{equation}
	\label{eq:fedavg}
	\theta \gets \sum_{i \in \mathcal{CS}} w_i \theta_i,
\end{equation}
where $w_i = \frac{N_i}{\sum_{k \in \mathcal{CS}} N_k}$.

\subsubsection{Hedonic Coalition Formation for Federated Learning Client Selection (\(\mathcal{CS}\))}

ICBAC formulates FL client selection as a hedonic coalition formation problem in which channels form coalitions based on declared preferences over collaboration partners. We adopt the SCC-based mechanism in \cite{Flip2023}, which provides core stability and strategy-proofness under friend-oriented preferences. Unlike similarity-based selection that may require exposing sensitive criteria or distributions, our mechanism only requires each agent to declare a list of ``friends,'' while reasons remain private.

\paragraph{\textbf{Theoretical Foundation:}}
Other client selection mechanisms may be unstable when participants have incentives to misreport preferences \cite{Flip2023,Liu2025,chen2025}. The SCC mechanism mitigates strategic manipulation by ensuring truthful reporting is dominant.

\begin{definition}[Hedonic Coalition Formation]
	A hedonic coalition formation problem consists of a finite set of agents $N = \{1, 2, \ldots, n\}$, where each agent $AI_i \in N$ has preferences $\succeq_{AI_i}$ over coalitions containing $AI_i$. An outcome is a partition $\pi$ of $N$, where $\pi_{AI_i}$ denotes the coalition containing agent $AI_i$.
\end{definition}

\begin{definition}[Core Stability]
	A partition $\pi$ is core stable if there exists no coalition $S \subseteq N$ such that for all $AI_i \in S$, $S \succ_{AI_i} \pi_{AI_i}$.
\end{definition}

\begin{definition}[Strategy-Proofness]
	A mechanism $\phi$ is strategy-proof if no agent can benefit by misreporting preferences, i.e., $\phi_{AI_i}(\succeq) \succeq_{AI_i} \phi_{AI_i}(\succeq'_{AI_i}, \succeq_{-AI_i})$ for all $AI_i$.
\end{definition}

\paragraph{\textbf{Friend-Oriented Preferences in Federated Learning:}}
Each agent $AI_i$ declares a friend list $F(\succeq_{AI_i}) \subseteq N\backslash\{AI_i\}$; all non-friends are treated as enemies. This preserves privacy because the criteria behind preferences are not disclosed.

\begin{definition}[Friend-Oriented Preferences]
	Agent $AI_i$'s preferences $\succeq_{AI_i}$ are friend-oriented if they can be represented by a partition of $N\backslash\{AI_i\}$ into friends $F(\succeq_{AI_i})$ and enemies $E(\succeq_{AI_i})$ such that:
	\begin{itemize}
		\item \textbf{(E) Enemy Aversion}: for each coalition $S$ and each enemy $e \notin S$, $S \succ_{AI_i} S \cup \{e\}$.
		\item \textbf{(F) Friend Preference}: for each coalition $S$, each friend $f \notin S$, and any set of enemies $E \subseteq E(\succeq_{AI_i}) \backslash S$, $S \cup \{f\} \cup E \succ_{AI_i} S$.
	\end{itemize}
\end{definition}

\paragraph{\textbf{SCC-Based Coalition Formation Algorithm:}}

\begin{figure*}[!h]
	\centering
	\includegraphics[scale=0.6]{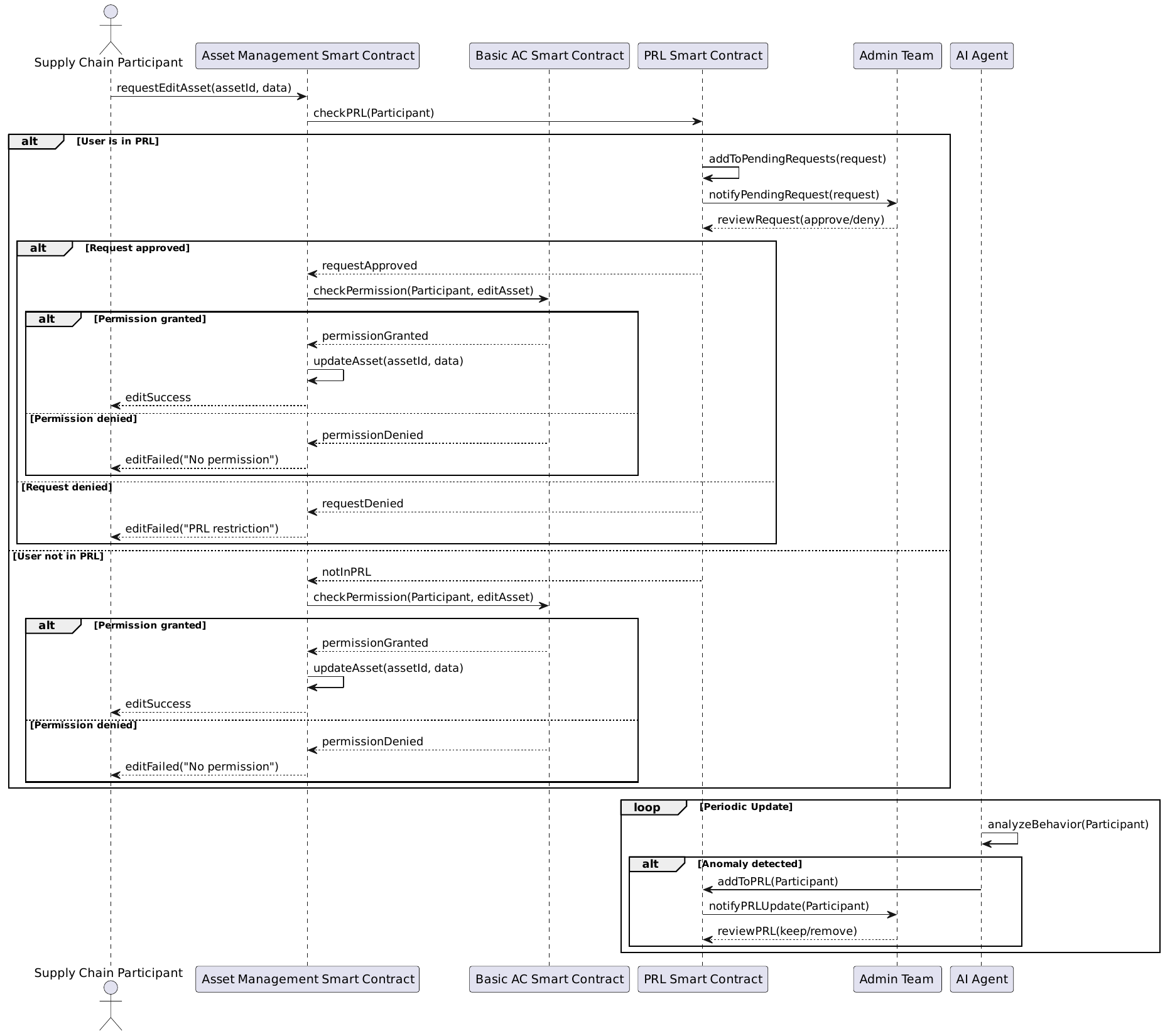}
	\caption{UML diagram of the overall workflow of asset updates and access control within a supply chain channel.}
	\label{fig:uml}
\end{figure*}

We construct the friendship graph $\Gamma(\succeq) = (N, A)$ where $A = \{(AI_i, AI_j) : AI_j \in F(\succeq_{AI_i})\}$ and compute SCCs using Tarjan’s algorithm \cite{Tarjan2024}. The SCC partition yields core stability and strategy-proofness guarantees \cite{Flip2023}.

\begin{theorem}[SCC Partition Properties]
	For any friend-oriented federated learning problem, the SCC partition $\pi^{SCC}$ satisfies: (i) core stability, (ii) group strategy-proofness, and (iii) computability in $O(|N|+|A|)$ time via Tarjan’s algorithm.
\end{theorem}

\begin{algorithm}[t]
	\caption{Coalition Formation for FL Client Selection}
	\label{alg:cf}
	\begin{algorithmic}[1]
		\REQUIRE Set of AI agents $N = \{AI_1, \ldots, AI_n\}$ and friend lists $\{F(\succeq_{AI_i})\}_{i=1}^n$
		\ENSURE Coalition partition $\pi^{SCC} = \{V_1, \ldots, V_k\}$
		\STATE Construct $\Gamma(\succeq) = (N, A)$ with edges $(AI_i, AI_j)$ if $AI_j \in F(\succeq_{AI_i})$
		\STATE Compute SCCs using Tarjan’s algorithm: $SCCs \leftarrow TarjanSCC(\Gamma(\succeq))$
		\STATE $\pi^{SCC} \leftarrow \{V_1, \ldots, V_k\}$ where each $V_\ell$ is the vertex set of SCC $\ell$
		\FOR{each coalition $V_\ell \in \pi^{SCC}$}
		\IF{$|V_\ell| \geq minCoalitionSize$}
		\STATE Initialize FL for $V_\ell$
		\ELSE
		\STATE Treat $V_\ell$ as local-only training (no FL)
		\ENDIF
		\ENDFOR
		\STATE \textbf{return} $\pi^{SCC}$
	\end{algorithmic}
\end{algorithm}

\subsection{Overall Workflow} \label{ss:ow}

Figure~\ref{fig:uml} summarizes the end-to-end workflow. Participants in a channel submit asset access/update requests. $\mathcal{SC}_{AM}$ executes the requested operation only after eligibility and permission checks via $\mathcal{SC}_{PRL}$ and $\mathcal{SC}_{BAC}$. In parallel, the channel AI agent monitors access behavior. If anomalous behavior is detected, the participant is added to $\mathcal{PRL}_i$ and subsequent requests are marked as \textit{Pending} until reviewed by the channel admin team $\mathcal{T}_i$. Admin teams may also promote or demote privileges by updating \textit{ExtraPrivilege}. Across channels, AI agents participate in coalition-based FL according to the client selection mechanism, improving anomaly detection without sharing raw data across supply chains.

\section{Evaluations and Experimental Results} \label{sec:eval}

This section evaluates ICBAC with respect to blockchain performance, anomaly detection effectiveness, federated learning behavior, and comparative system capabilities. The evaluation proceeds in three stages. First, the blockchain layer is benchmarked in terms of latency and throughput and compared against representative static blockchain-based access control models \cite{Li2023,Sarfaraz2023,Li2024}. Second, the AI layer is evaluated through federated anomaly detection under both IID and non-IID data distributions. Finally, ICBAC is compared against recent frameworks in terms of functional and architectural features.

All experiments were conducted on a system equipped with an Intel\textsuperscript{\textregistered} Core\texttrademark{} i7 processor, 16\,GB RAM, running Ubuntu~22.04.3~LTS. Federated learning was implemented in Python~3.12.6 using PyTorch~2.7.1. Smart contracts were developed in Go~1.22.2 and deployed on a local Hyperledger Fabric~2.5 network.

\subsection{Blockchain Evaluation}

We evaluate the blockchain layer of ICBAC on a Hyperledger Fabric network, focusing on transaction latency and throughput as key indicators for scalability and practical deployment in SCM. Latency is measured as the elapsed time between transaction submission and commitment (ms), and throughput is measured as committed transactions per second (TPS).

The evaluation is conducted in two phases. First, we analyze latency and throughput of core functions across the Asset Management, Basic Access Control, and Permission Revoke List smart contracts to assess internal scalability. Second, we compare ICBAC against MedTrace \cite{Li2023}, AccessChain \cite{Sarfaraz2023}, and ProChain \cite{Li2024} to quantify the performance cost of dynamic, AI-aware access control.

For fairness, all systems were implemented in Go and deployed using identical Hyperledger Fabric configurations consisting of four peer nodes, three orderer nodes, and two certificate authorities.

\begin{figure*}[!h]
	\centering
	\fontfamily{ptm}\selectfont 
	\hspace*{\fill}
	\begin{subfigure}{0.45\textwidth} 
		\centering
		\includegraphics[width=\textwidth]{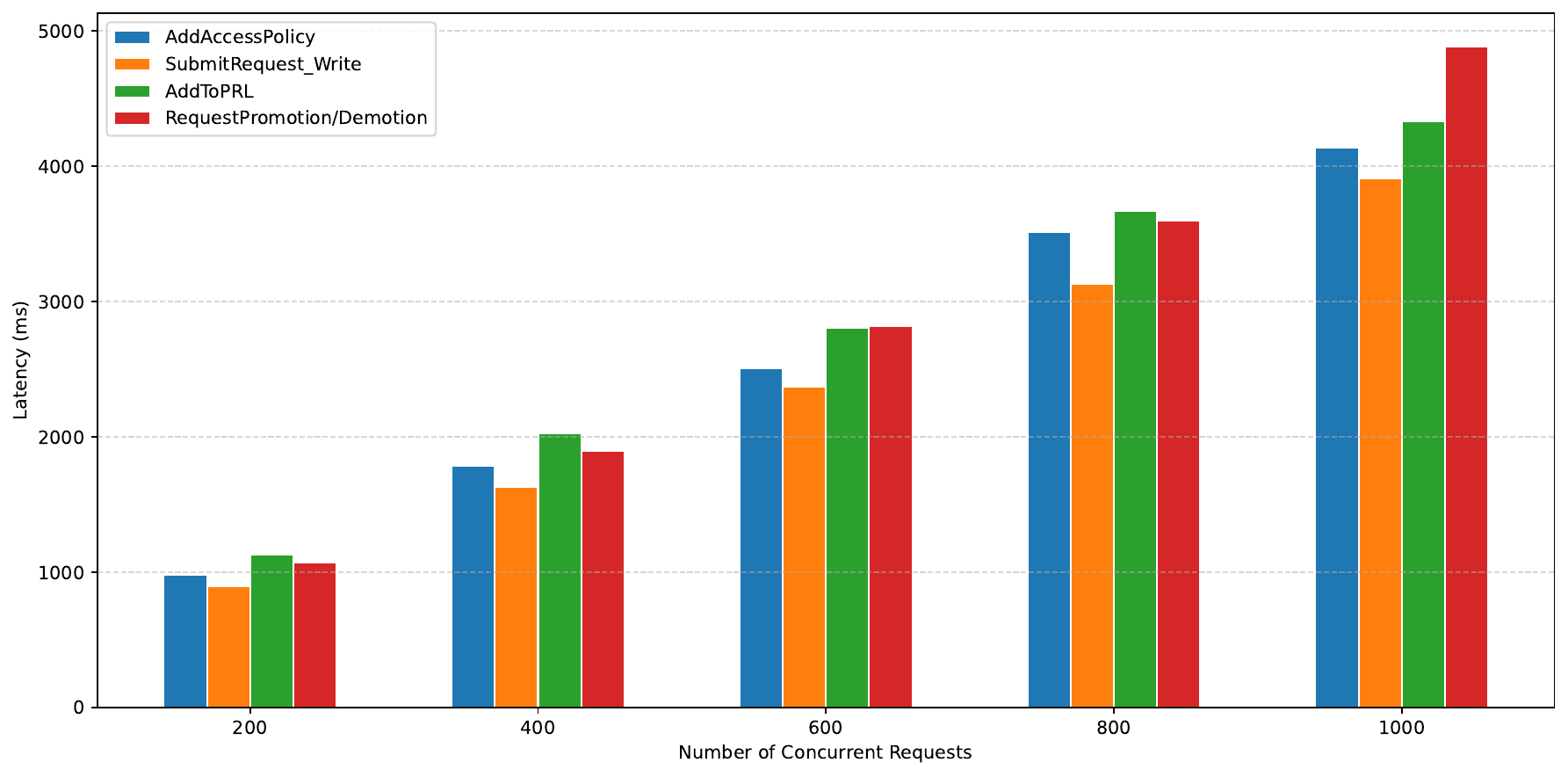}
		\caption{Latency of Write Requests}
		\label{fig:write_latency_plot}
	\end{subfigure}
	\hfill
	\begin{subfigure}{0.45\textwidth} 
		\centering
		\includegraphics[width=\textwidth]{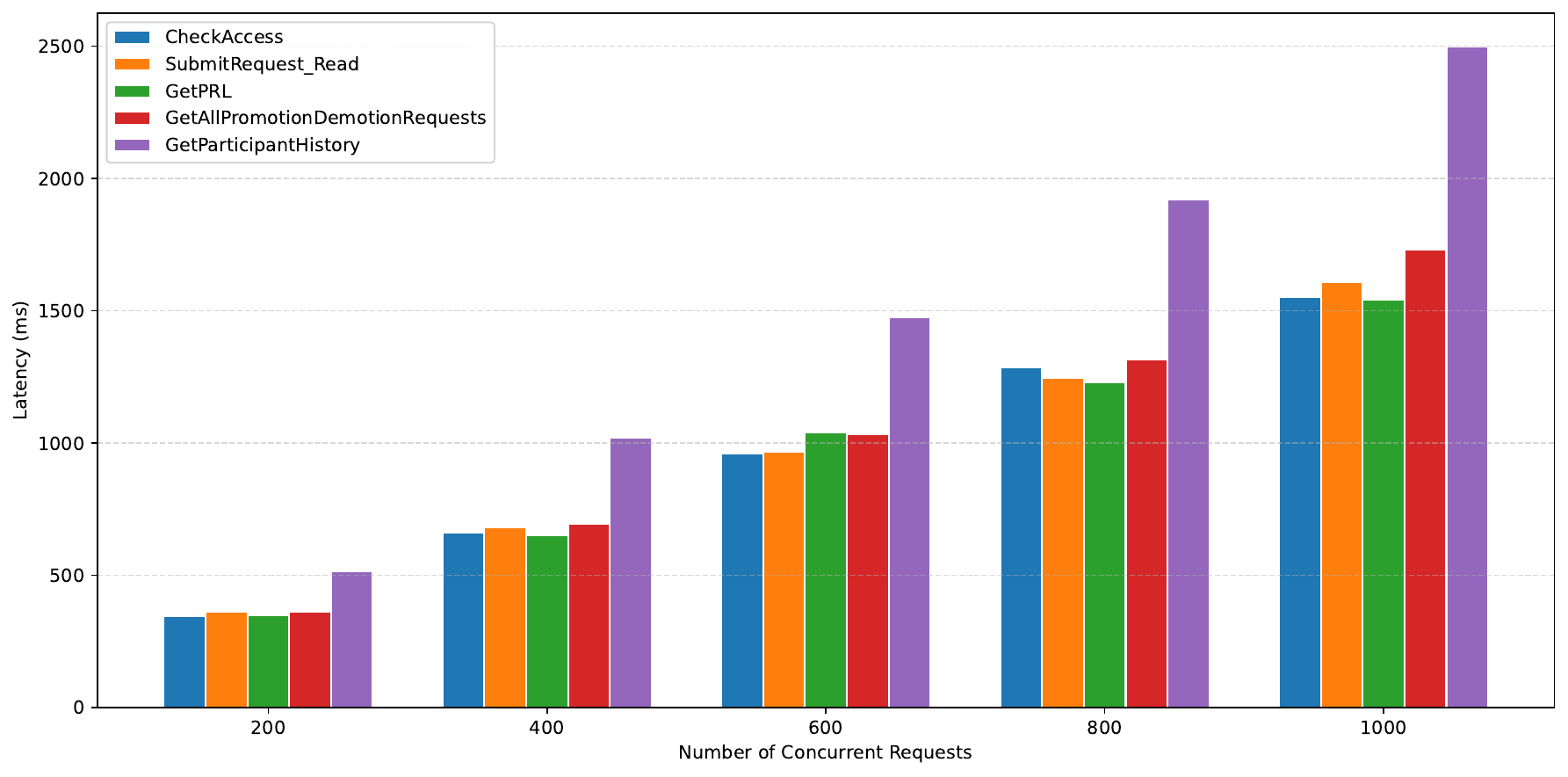}
		\caption{Latency of Read Requests}
		\label{fig:read_latency_plot}
	\end{subfigure}
	\hspace*{\fill}
	\par\vspace{1em} 
	\hspace*{\fill}
	\begin{subfigure}{0.45\textwidth} 
		\centering
		\includegraphics[width=\textwidth]{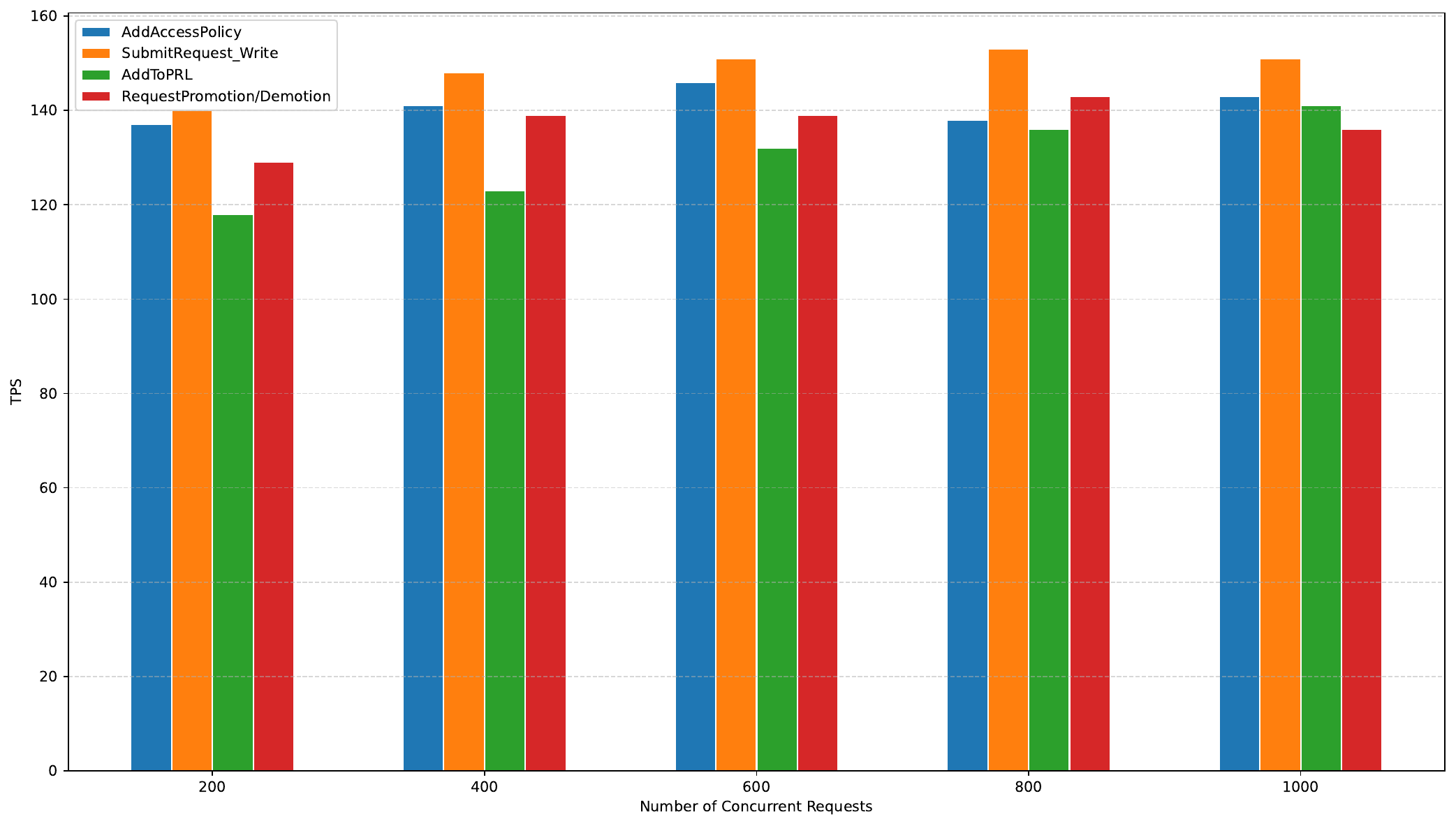}
		\caption{Throughput of Write Requests}
		\label{fig:write_throughput_plot}
	\end{subfigure}
	\hfill
	\begin{subfigure}{0.45\textwidth} 
		\centering
		\includegraphics[width=\textwidth]{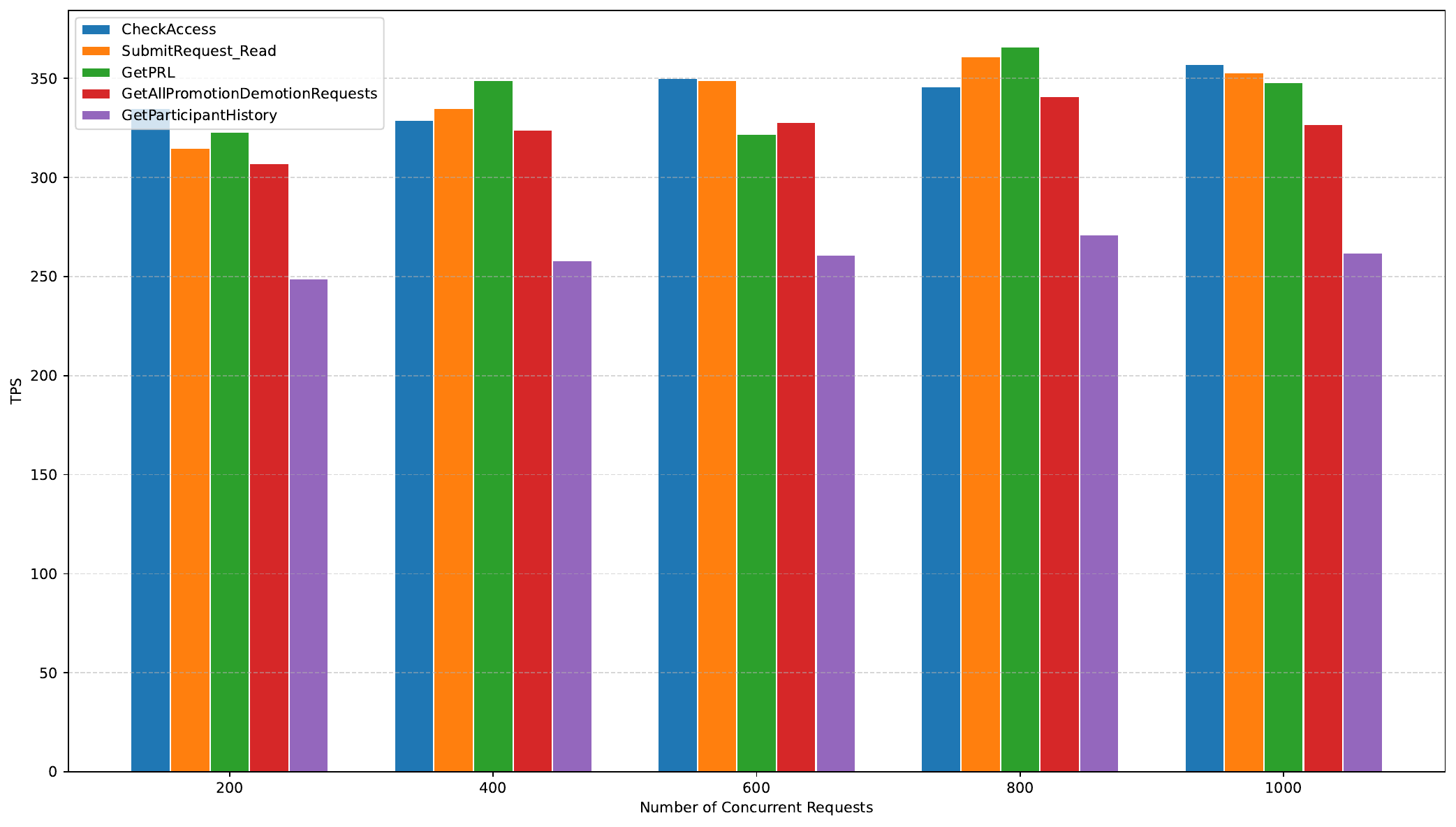}
		\caption{Throughput of Read Requests}
		\label{fig:read_throughput_plot}
	\end{subfigure}
	\hspace*{\fill} 
	\caption{Latency and throughput analysis of ICBAC smart contract functions.}
	\label{fig:our_bc_plot}
\end{figure*}

Figures~\ref{fig:write_latency_plot} and~\ref{fig:read_latency_plot} show that latency increases approximately linearly with concurrent requests. Write operations exhibit higher latency due to endorsement, ordering, and validation overhead, while read operations remain lower because they bypass consensus. Figures~\ref{fig:write_throughput_plot} and~\ref{fig:read_throughput_plot} show throughput increasing with load until saturation, after which it stabilizes; read throughput is higher than write throughput due to lower overhead.

\begin{figure*}[!h]
	\centering
	\fontfamily{ptm}\selectfont 
	\hspace*{\fill}
	\begin{subfigure}{0.35\textwidth} 
		\centering
		\includegraphics[width=\textwidth]{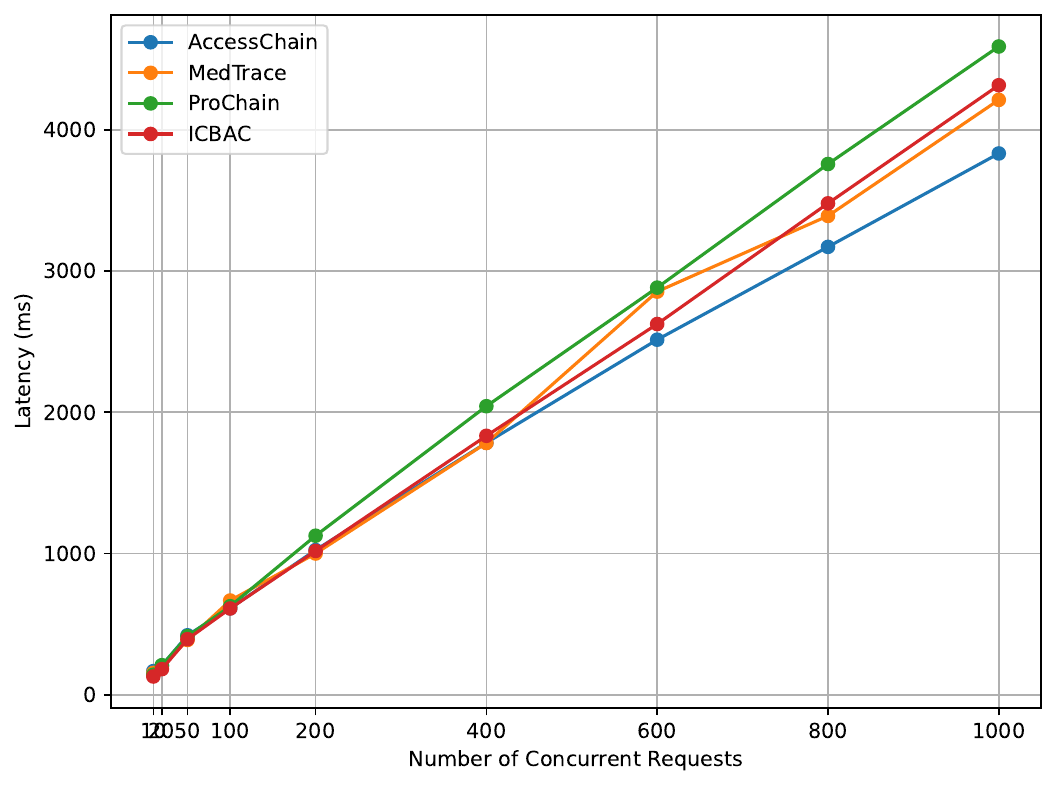}
		\caption{Latency of Write Requests}
		\label{fig:write_latency_comparison}
	\end{subfigure}
	\hfill
	\begin{subfigure}{0.35\textwidth} 
		\centering
		\includegraphics[width=\textwidth]{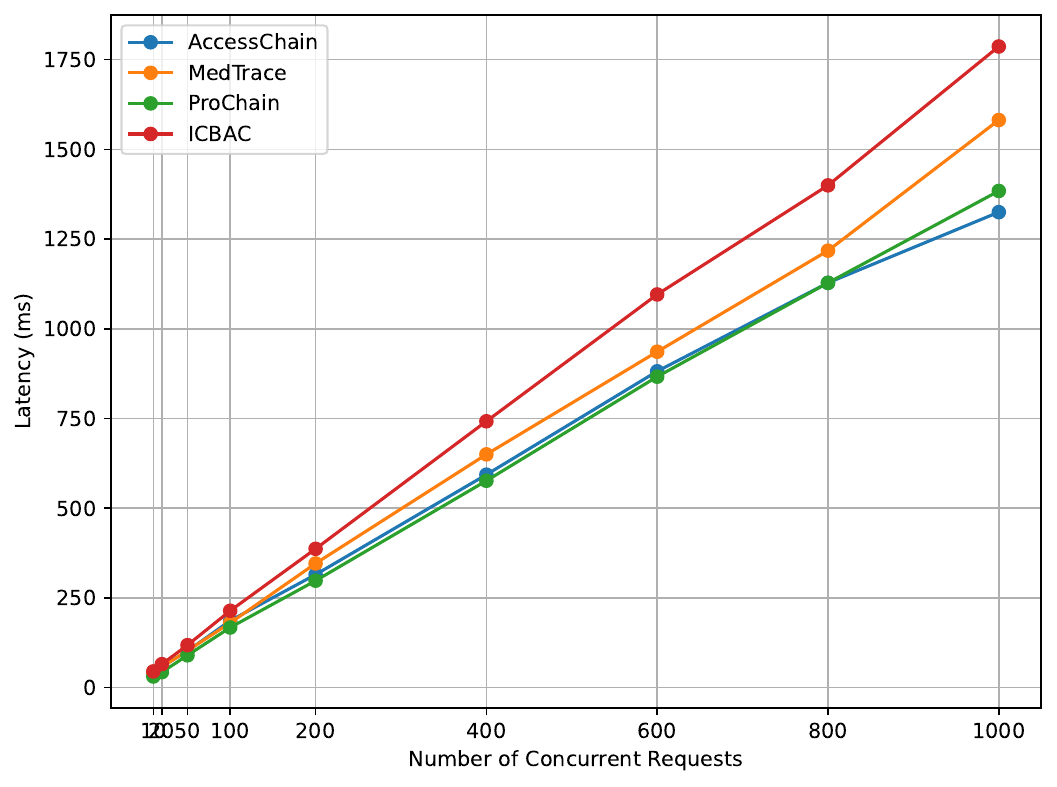}
		\caption{Latency of Read Requests}
		\label{fig:read_latency_comparison}
	\end{subfigure}
	\hspace*{\fill}
	\par\vspace{1em} 
	\hspace*{\fill}
	\begin{subfigure}{0.35\textwidth} 
		\centering
		\includegraphics[width=\textwidth]{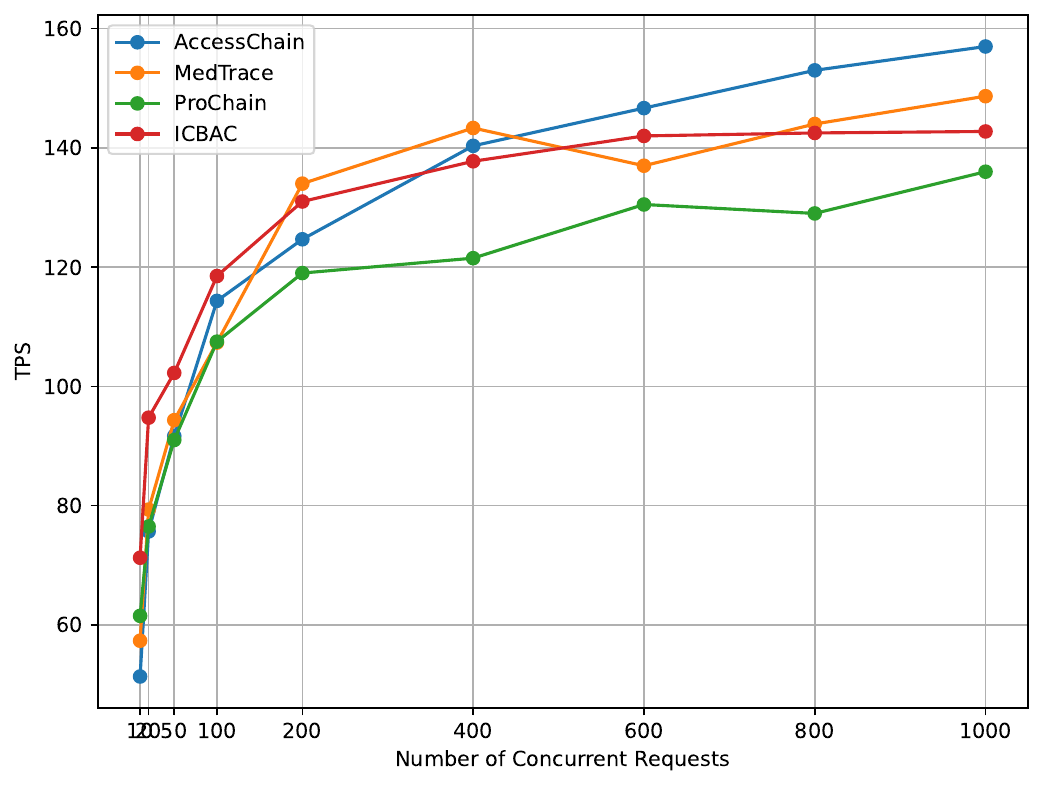}
		\caption{Throughput of Write Requests}
		\label{fig:write_throughput_comparison}
	\end{subfigure}
	\hfill
	\begin{subfigure}{0.35\textwidth} 
		\centering
		\includegraphics[width=\textwidth]{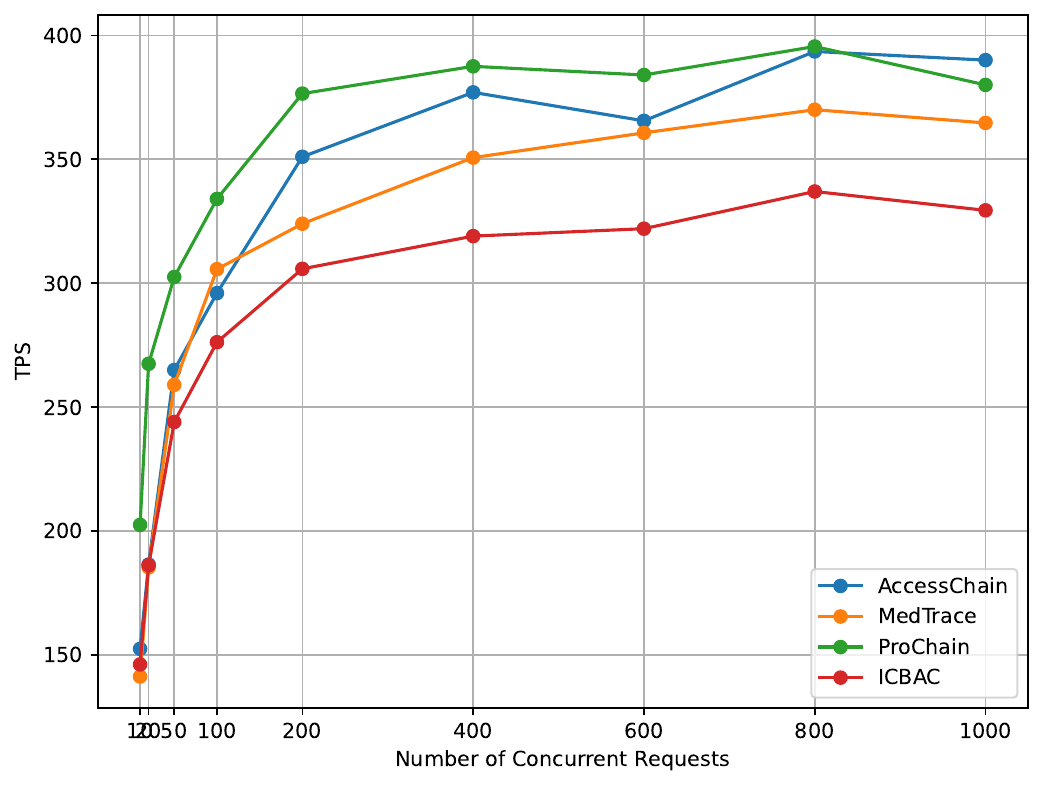}
		\caption{Throughput of Read Requests}
		\label{fig:read_throughput_comparison}
	\end{subfigure}
	\hspace*{\fill} 
	\caption{Comparative performance analysis of ICBAC against recent static blockchain-based access control models in SCM.}
	\label{fig:comparision_bc_plot}
\end{figure*}

Figure~\ref{fig:comparision_bc_plot} compares ICBAC with MedTrace \cite{Li2023}, AccessChain \cite{Sarfaraz2023}, and ProChain \cite{Li2024}. ICBAC achieves write latency and throughput comparable to all baselines. Read operations show slightly higher latency and lower throughput, consistent with the additional checks required for dynamic permissions and PRL verification.

To assess statistical significance, independent two-sample t-tests were conducted across read throughput, write throughput, read latency, and write latency using nine load levels (10--1000 concurrent requests), yielding 16 degrees of freedom. The pooled-variance t-statistic was computed as:
\[
t = \frac{\mu_1 - \mu_2}{s_p \sqrt{1/n_1 + 1/n_2}}.
\]
Results indicate that 11 out of 12 comparisons show no statistically significant difference ($p > 0.10$). A marginal difference is observed for read throughput when comparing ICBAC with ProChain ($t=-2.088$, $0.05 < p < 0.10$), corresponding to an 18.6\% reduction. This overhead stems from dynamic permission and anomaly checks, which are necessary for behavior-aware access control in SCM.

\subsection{Federated Learning Evaluation}

This subsection evaluates the AI layer of ICBAC, focusing on federated anomaly detection performance and the impact of coalition-based client selection. Experiments are conducted under both IID and non-IID data distributions to reflect realistic supply chain heterogeneity.

\subsubsection{Dataset, Preprocessing, and Evaluation Parameters}

We use the DataCoSupplyChain dataset \cite{DataCo}, which contains over 180{,}000 real-world supply chain transaction records across 23 geographic regions. Each region is modeled as an independent blockchain channel and FL client, and each channel includes three participants: factory, distributor, and retailer.

Raw records are transformed into participant-specific behavioral logs. Approximately 10\% of records are injected with controlled anomalies. Both IID and non-IID datasets are generated by selectively applying anomaly scenarios---such as abnormal shipping delays, inflated production reports, unauthorized resource usage, and manipulated sales---so that identical behaviors may be anomalous in some channels but normal in others.

The following evaluation metrics are used:
\begin{itemize}
	\item \textbf{Accuracy}:
	\[
	\frac{\mathrm{TP} + \mathrm{TN}}{\mathrm{TP} + \mathrm{TN} + \mathrm{FP} + \mathrm{FN}}
	\]
	\item \textbf{Precision}:
	\[
	\frac{\mathrm{TP}}{\mathrm{TP} + \mathrm{FP}}
	\]
	\item \textbf{Recall}:
	\[
	\frac{\mathrm{TP}}{\mathrm{TP} + \mathrm{FN}}
	\]
	\item \textbf{F1-score}:
	\[
	2 \cdot \frac{\text{Precision} \cdot \text{Recall}}{\text{Precision} + \text{Recall}}
	\]
	\item \textbf{Loss}: average cross-entropy loss per FL round
	\item \textbf{Privacy Leakage}:
	\[
	\frac{\sum \text{Shared Model Update Volume}}{\sum \text{Total Local Data Volume}}
	\]
	This metric is used as a communication-based exposure proxy rather than a formal privacy guarantee.
\end{itemize}

For IID data, models are trained with a fixed learning rate of 0.01 and one epoch per FL round. For non-IID data, the learning rate decays from 0.01 to 0.0004 with ten local epochs per round.

\begin{figure*}[!h]
	\centering
	\fontfamily{ptm}\selectfont 
	\begin{subfigure}{0.33\textwidth} 
		\centering
		\includegraphics[width=\textwidth]{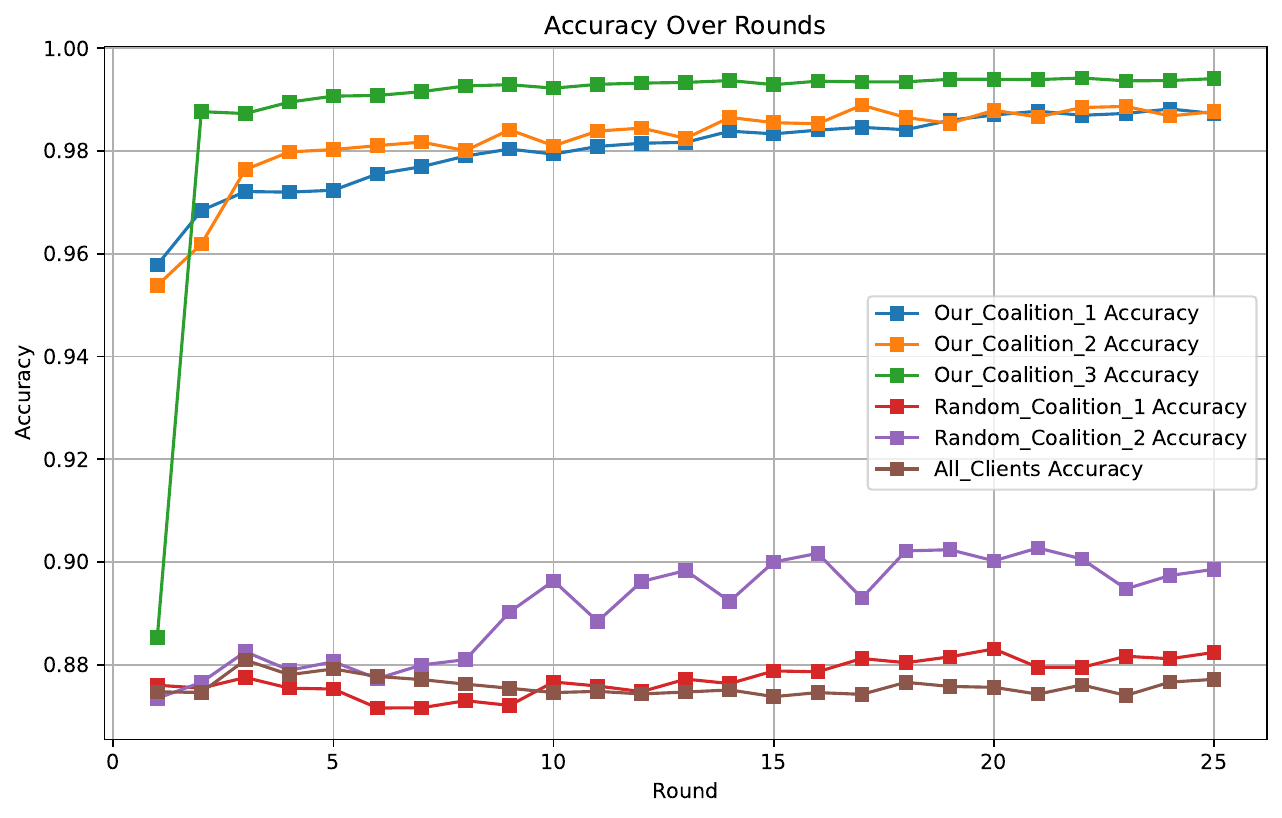}
		\caption{Accuracy}
		\label{fig:iid_accuracy}
	\end{subfigure}
	\hfill
	\begin{subfigure}{0.33\textwidth} 
		\centering
		\includegraphics[width=\textwidth]{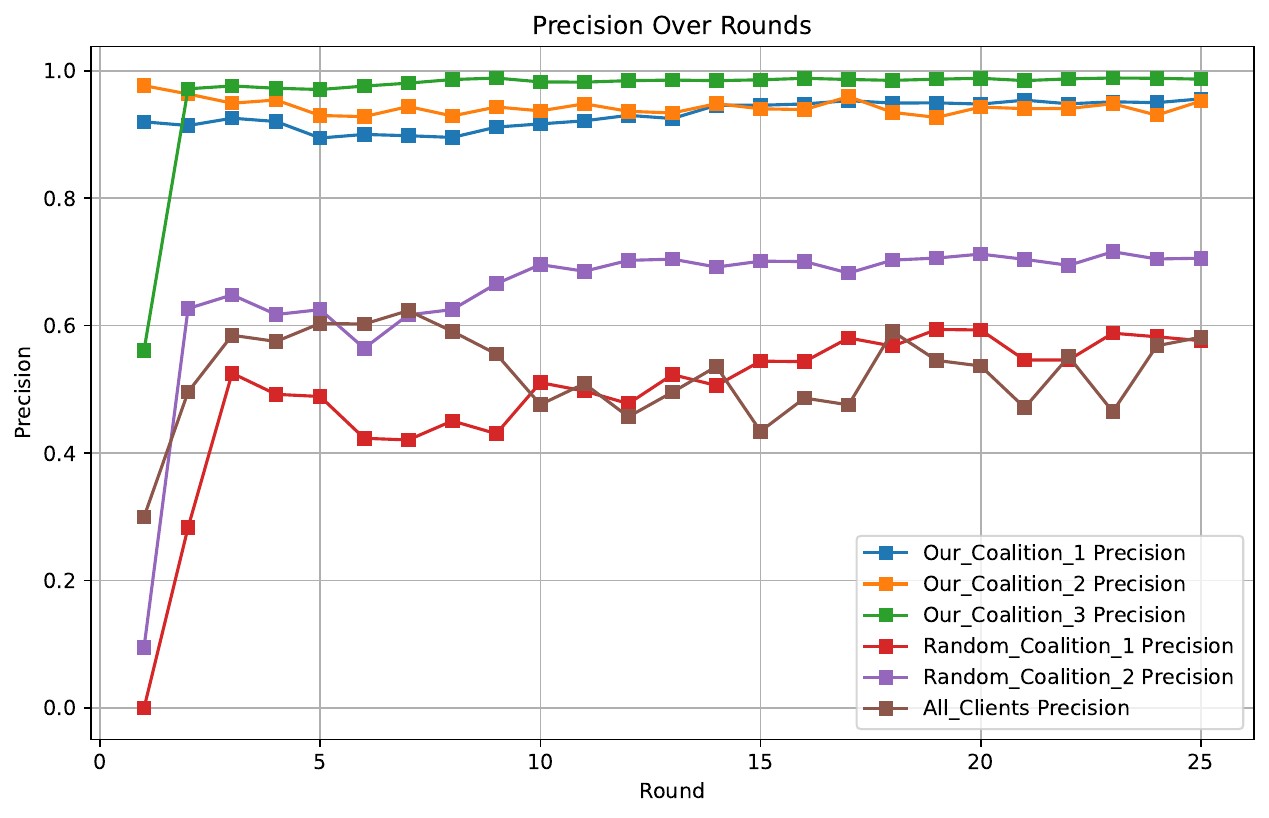}
		\caption{Precision}
		\label{fig:iid_precision}
	\end{subfigure}
	\hfill
	\begin{subfigure}{0.33\textwidth} 
		\centering
		\includegraphics[width=\textwidth]{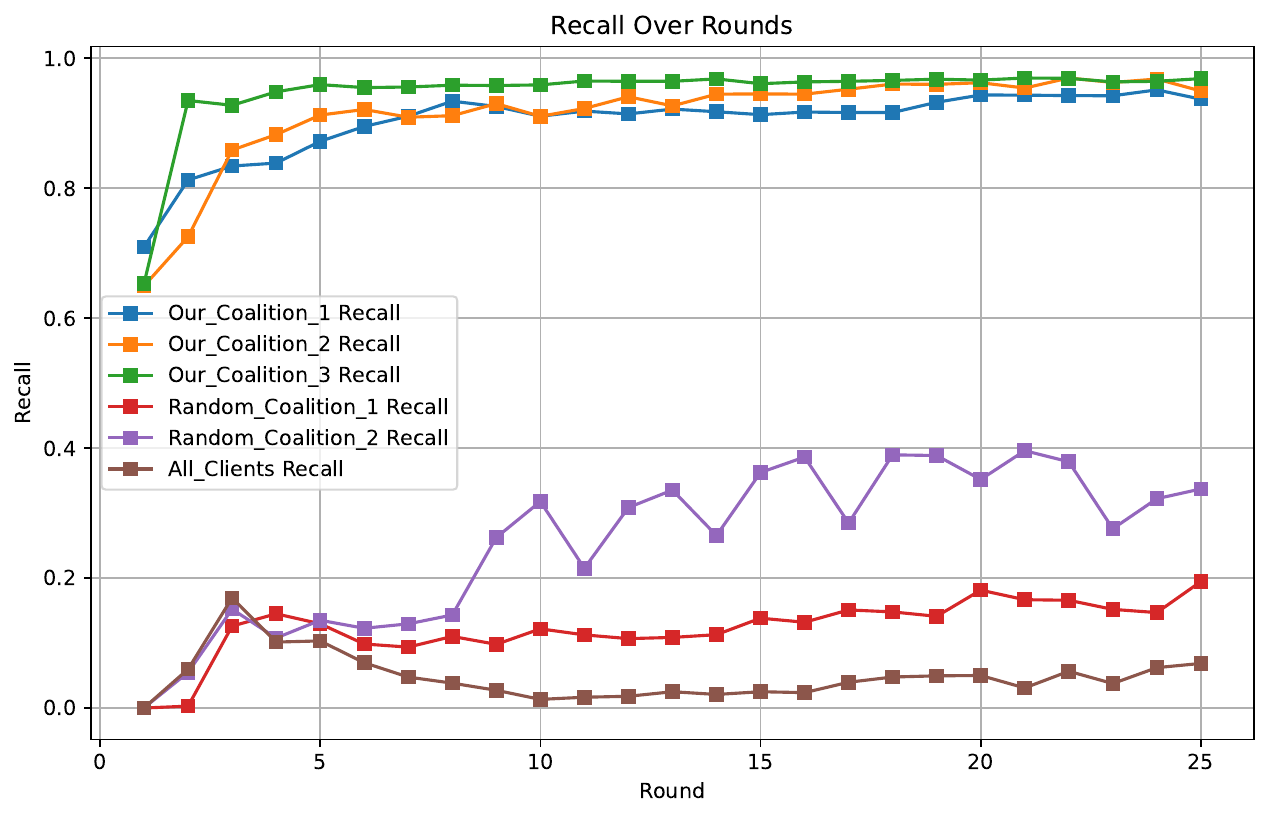}
		\caption{Recall}
		\label{fig:iid_recall}
	\end{subfigure}
	\par\vspace{1em} 
	\hspace*{\fill}
	\begin{subfigure}{0.33\textwidth} 
		\centering
		\includegraphics[width=\textwidth]{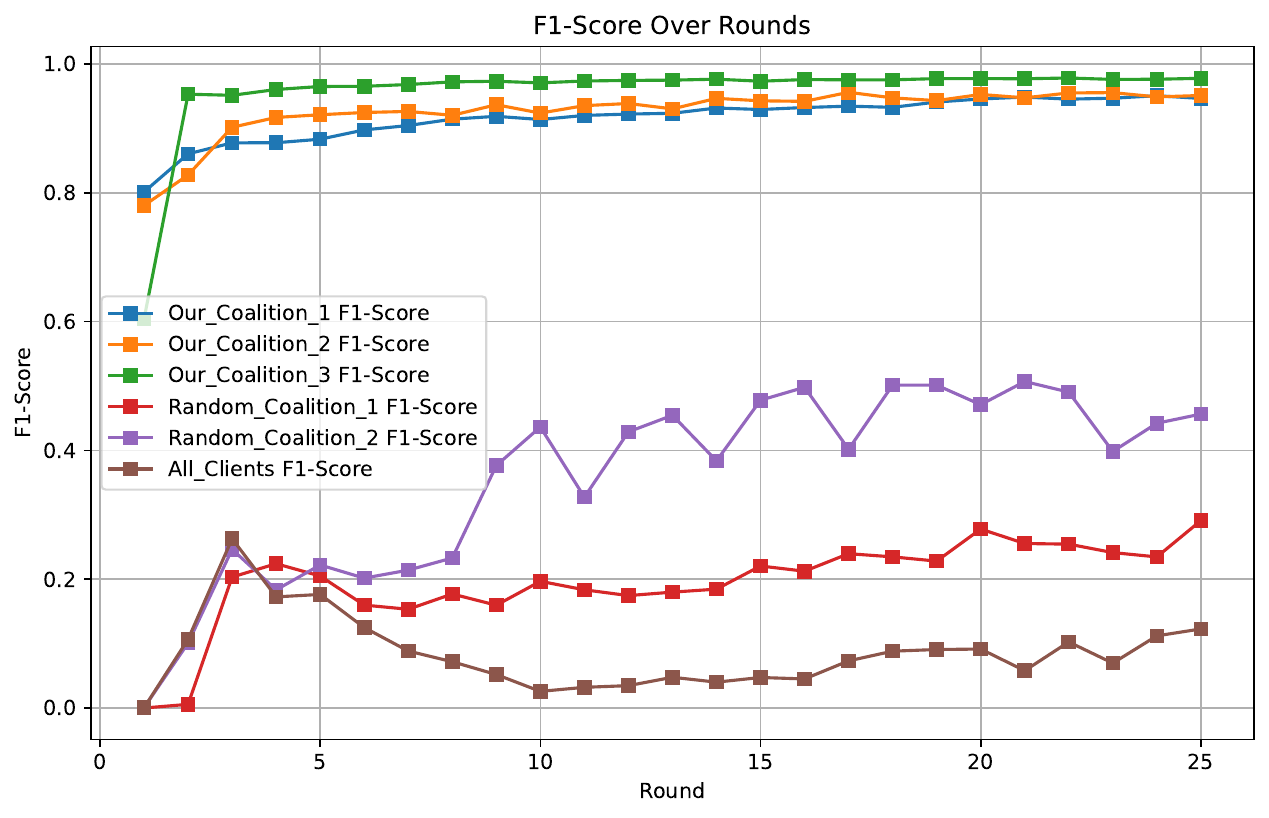}
		\caption{F1-score}
		\label{fig:iid_f1_score}
	\end{subfigure}
	\hfill
	\begin{subfigure}{0.33\textwidth} 
		\centering
		\includegraphics[width=\textwidth]{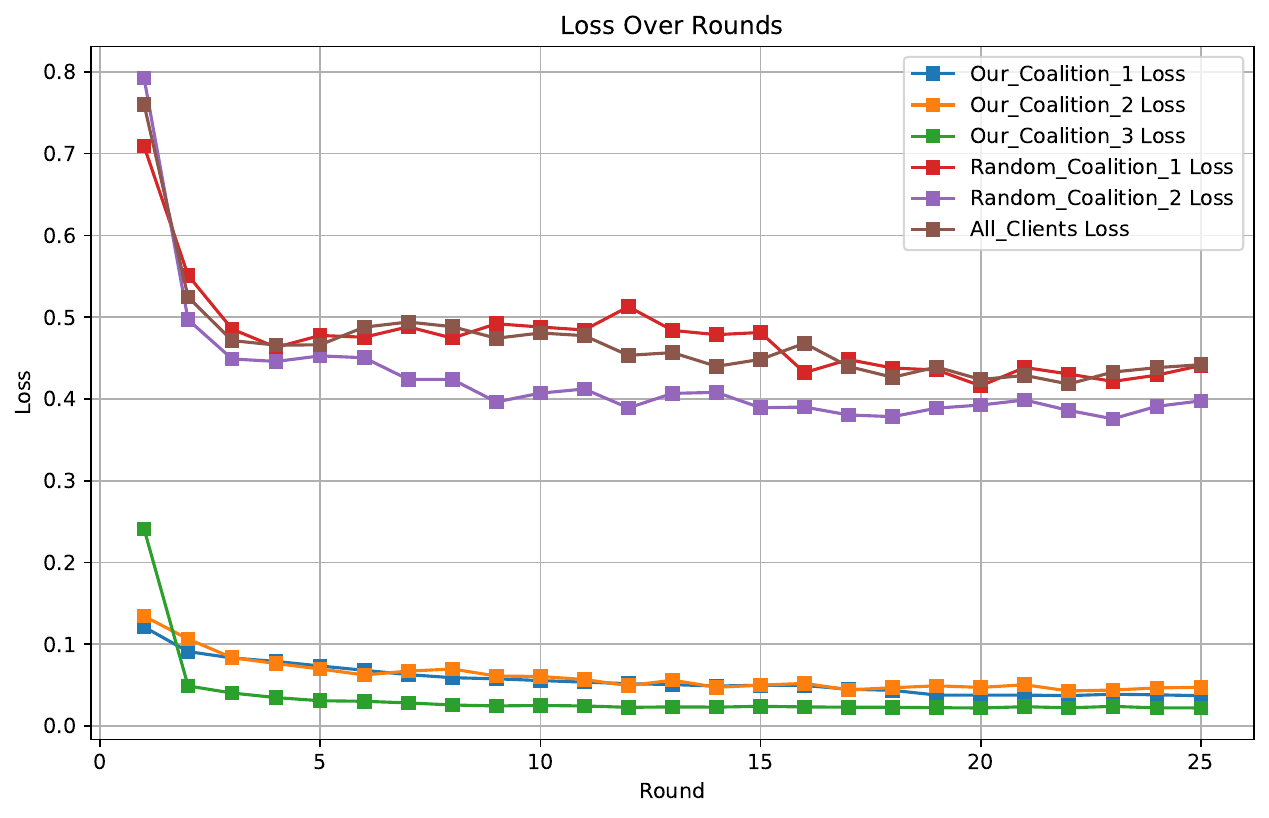}
		\caption{Loss}
		\label{fig:iid_loss}
	\end{subfigure}
	\hspace*{\fill} 
	\caption{Performance metrics for IID dataset across different coalitions.}
	\label{fig:iid_metrics}
\end{figure*}

\begin{figure*}[!h]
	\centering
	\fontfamily{ptm}\selectfont
	\begin{subfigure}{0.33\textwidth} 
		\centering
		\includegraphics[width=\textwidth]{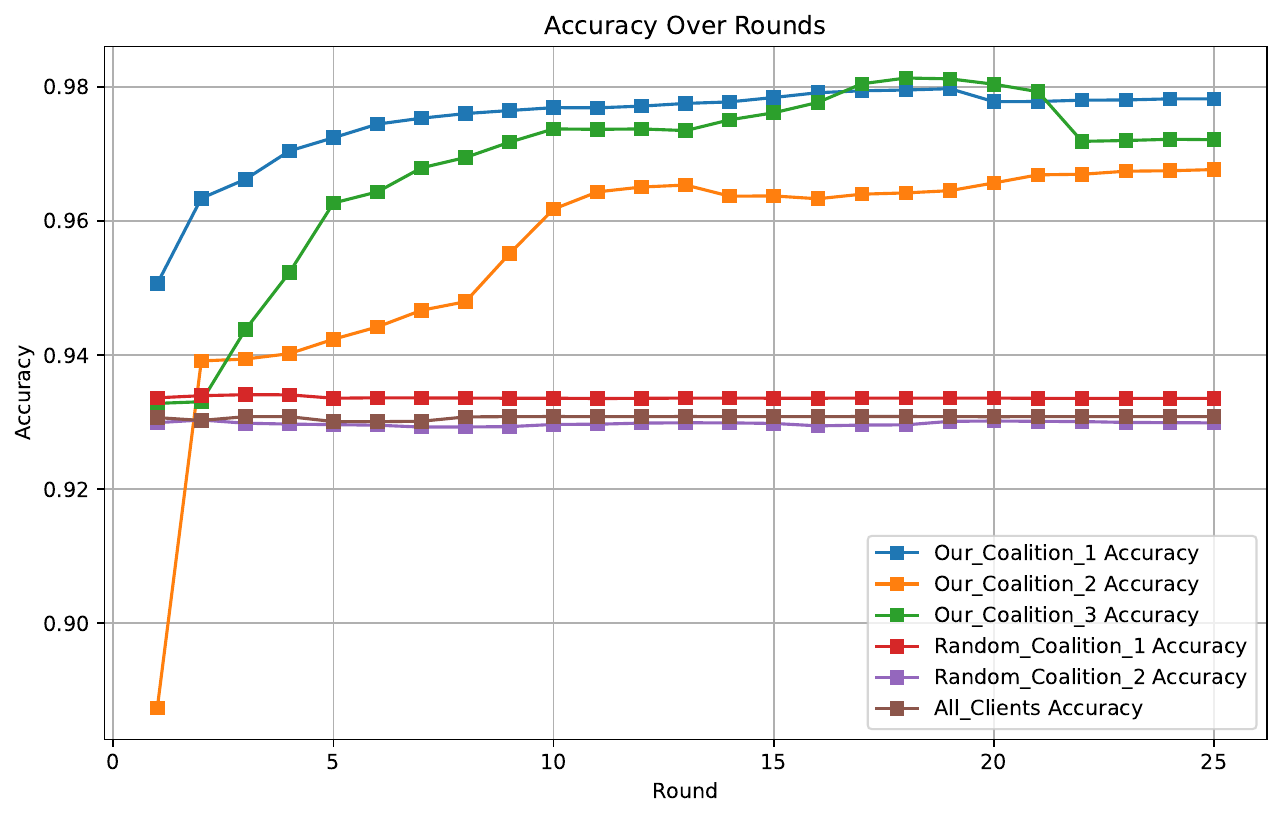}
		\caption{Accuracy}
		\label{fig:non_iid_accuracy}
	\end{subfigure}
	\hfill
	\begin{subfigure}{0.33\textwidth}
		\centering
		\includegraphics[width=\textwidth]{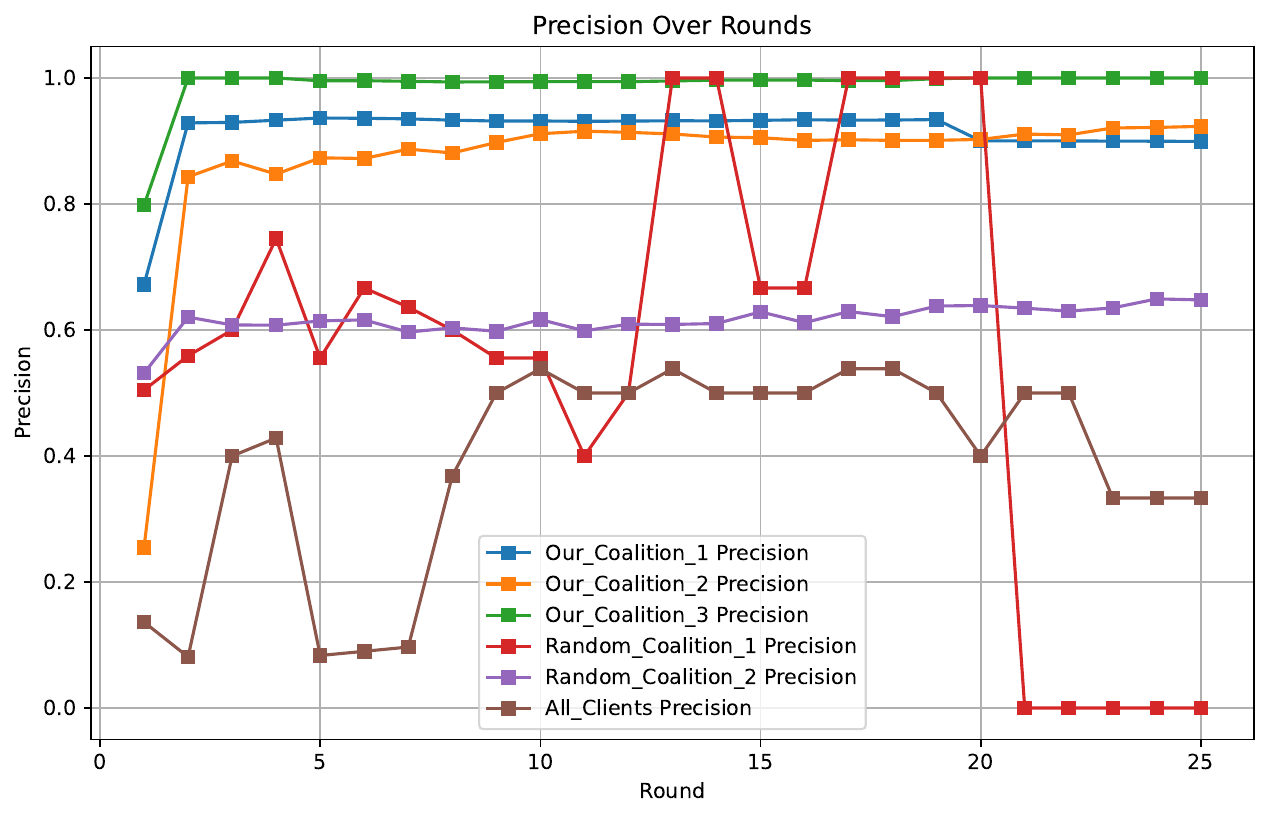}
		\caption{Precision}
		\label{fig:non_iid_precision}
	\end{subfigure}
	\hfill
	\begin{subfigure}{0.33\textwidth} 
		\centering
		\includegraphics[width=\textwidth]{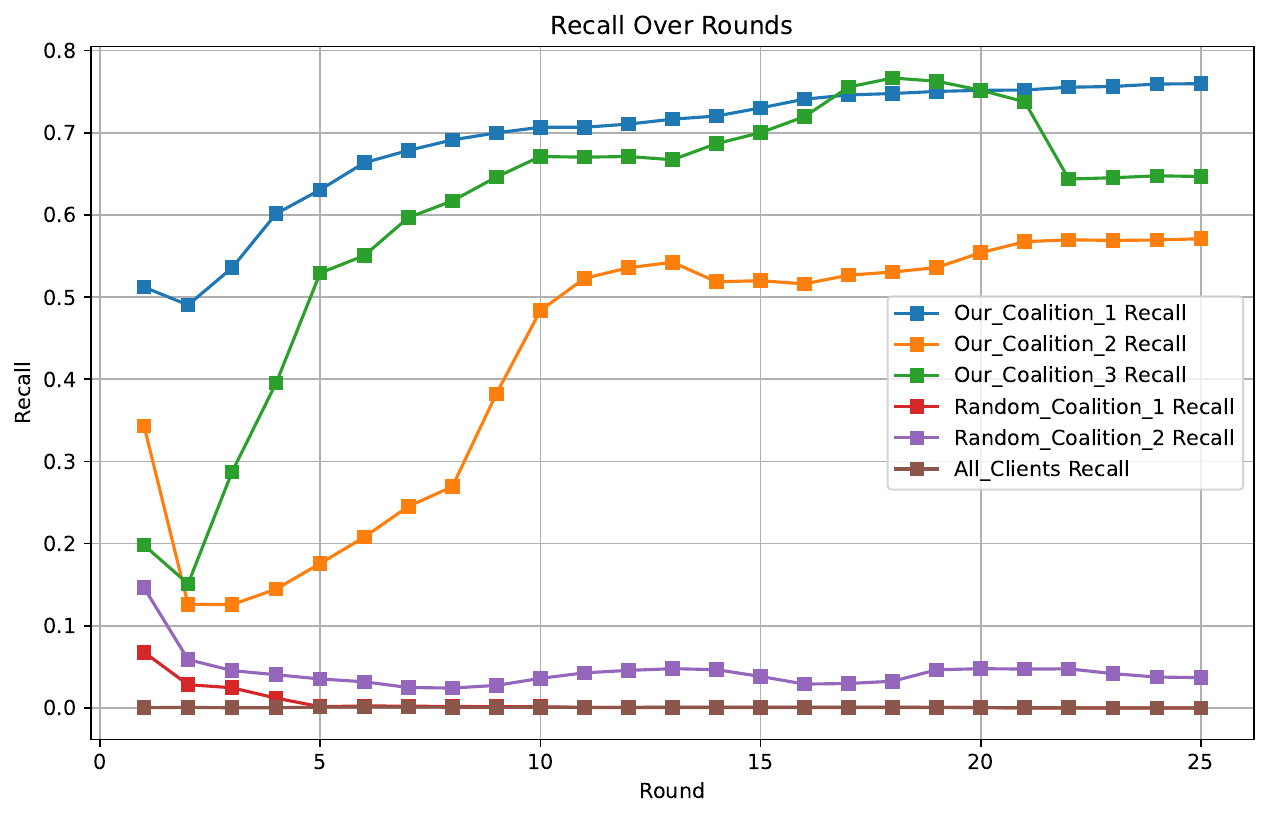}
		\caption{Recall}
		\label{fig:non_iid_recall}
	\end{subfigure}
	\par\vspace{1em}
	\hspace*{\fill}
	\begin{subfigure}{0.33\textwidth}
		\centering
		\includegraphics[width=\textwidth]{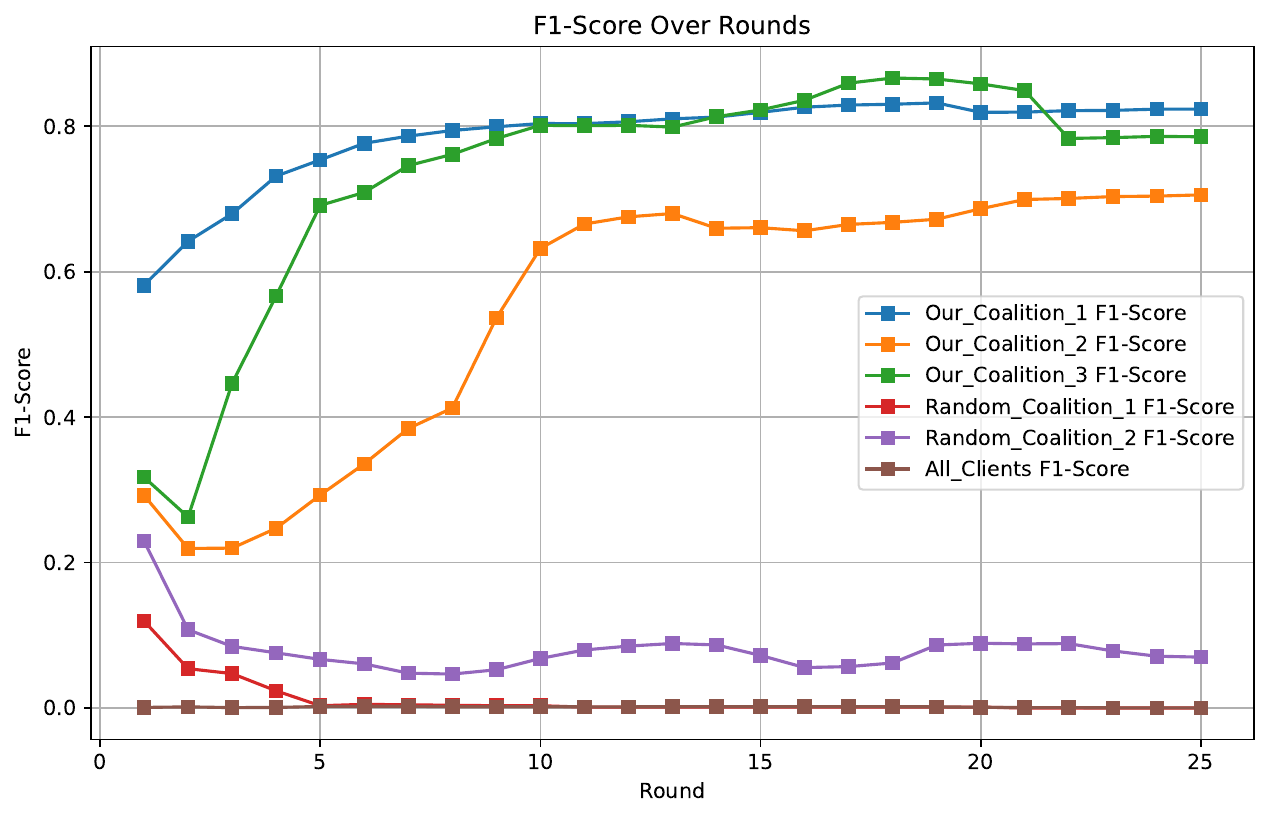}
		\caption{F1-score}
		\label{fig:non_iid_f1_score}
	\end{subfigure}
	\hfill
	\begin{subfigure}{0.33\textwidth} 
		\centering
		\includegraphics[width=\textwidth]{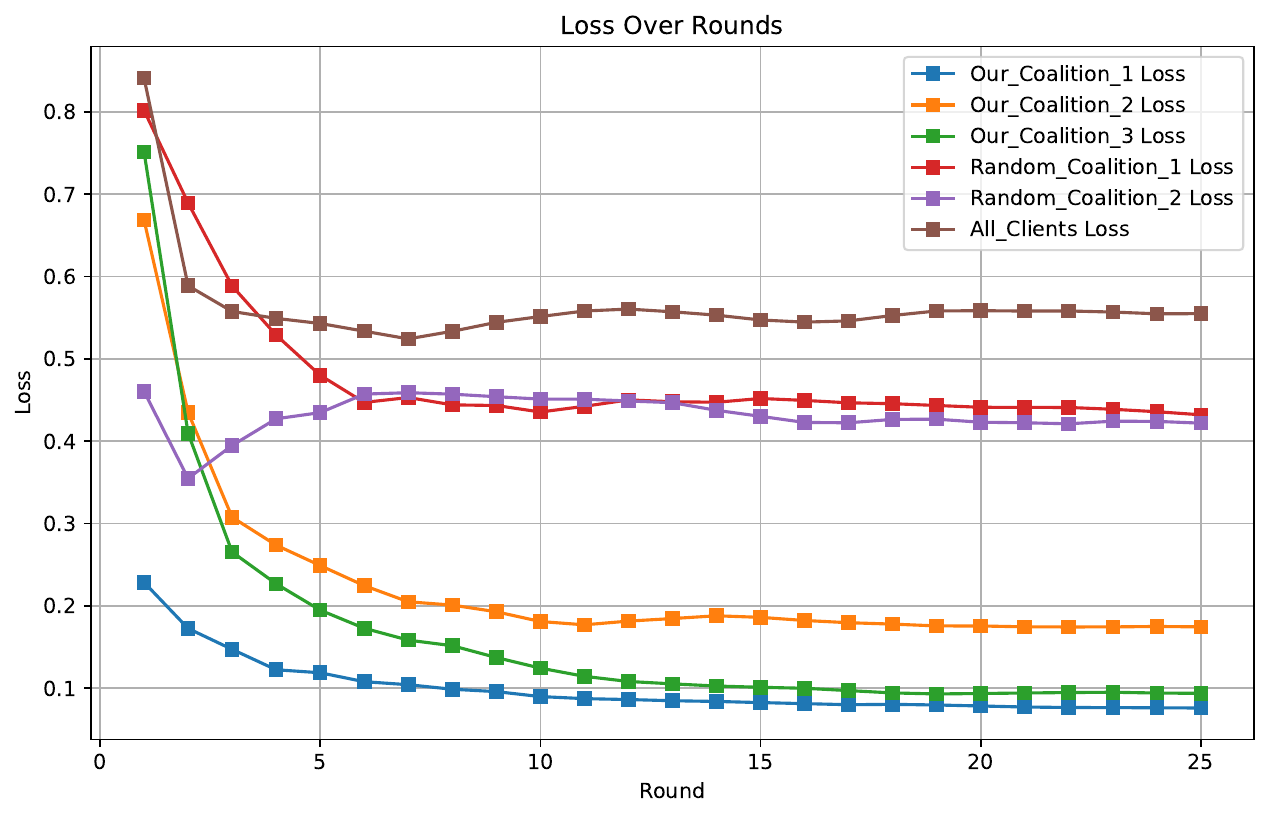}
		\caption{Loss}
		\label{fig:non_iid_loss}
	\end{subfigure}
	\hspace*{\fill}
	\caption{Performance metrics for Non-IID dataset across different coalitions.}
	\label{fig:non_iid_metrics}
\end{figure*}

Figures~\ref{fig:iid_metrics} and~\ref{fig:non_iid_metrics} summarize anomaly detection performance across coalitions and data regimes.

\subsubsection{Coalition Formation}
\label{ssec:coalition_formation}

We evaluate federated learning in ICBAC using three preference-based coalitions formed by the proposed hedonic coalition mechanism and compare them with two randomly formed coalitions and one all-inclusive coalition. The goal is to assess whether coalition-based participation improves convergence and anomaly detection performance under both IID and Non-IID data distributions.

\paragraph{IID vs Non-IID settings.}
In the IID setting, we assume that client data distributions are sufficiently similar such that training samples across channels represent comparable operational processes and anomaly signatures. In contrast, the Non-IID setting models realistic supply chain heterogeneity, where channels differ in transaction patterns, feature distributions, and anomaly semantics. This heterogeneity is intentionally reflected in our anomaly injection design: identical behaviors (e.g., a delay pattern or production inflation) may be labeled anomalous in some channels while being treated as normal in others, capturing the contextual nature of anomalous behavior in supply chains.

\paragraph{Why Non-IID federated learning typically underperforms.}
Under Non-IID distributions, federated learning faces fundamental optimization challenges, often referred to as \emph{client drift}. Each client minimizes a local objective induced by its own data distribution, and when these objectives are misaligned, local updates may point in different directions in parameter space. As a result, naive aggregation (e.g., FedAvg) can yield a global model that converges more slowly or settles at a suboptimal solution. In anomaly detection, this effect is amplified because (i) the boundary between normal and anomalous behavior is context-dependent, and (ii) the rarity and diversity of anomalous samples increases gradient variance. Consequently, the aggregated model may be pulled toward behaviors that are ``normal'' in some channels but anomalous in others, reducing recall or precision in heterogeneous settings.

\paragraph{Role of coalition formation in mitigating heterogeneity.}
ICBAC addresses this challenge by forming coalitions among behaviorally compatible channels. Instead of forcing all channels into a single federation, coalition formation restricts aggregation to clients that declare mutual compatibility, which acts as a practical proxy for distributional similarity without requiring channels to reveal sensitive criteria or raw statistics. This design reduces the degree of objective misalignment across participating clients and mitigates client drift, thereby improving convergence stability and anomaly detection performance relative to random or all-inclusive aggregation.

\paragraph{Observed effects in experiments.}
Figures~\ref{fig:iid_metrics} and~\ref{fig:non_iid_metrics} report results for the IID and Non-IID settings, respectively. Across both regimes, preference-based coalitions converge faster and achieve higher F1-scores with lower loss compared to random coalitions and the all-inclusive coalition. The performance gap is more pronounced in the Non-IID regime, where unrestricted aggregation is more strongly affected by distribution mismatch and contextual anomaly semantics. While Non-IID results are generally lower than IID results (as expected from the above discussion), the coalition mechanism consistently improves robustness and reduces degradation by limiting aggregation to more compatible channels.

\paragraph{One-time coalition formation.}
Coalition formation in our experiments is executed once during initialization based on declared preferences, rather than being recomputed in every FL round. This is motivated by the nature of the selection criterion: we use supply chain context compatibility (behavioral similarity and collaboration preference) as the primary driver, which changes infrequently compared to transient factors such as device availability. Recomputing coalitions at every round would therefore add overhead without providing meaningful benefits. Channels may update their friend lists and re-enter the matching pool when their collaboration preferences change.

\subsubsection{Comparison with Central Model}

\begin{table*}[!h]
	\centering
	\fontfamily{ptm}\selectfont
	\caption{Comparison of performance metrics for proposed FL-based models and central models within coalitions.}
	\label{tab:performance_metrics}
	\resizebox{1.1\textwidth}{!}{
	\begin{tabular}{l cccc cccc c}
		\toprule
		& \multicolumn{4}{c}{\textbf{\fontfamily{ptm}\selectfont IID Dataset}} & \multicolumn{4}{c}{\textbf{\fontfamily{ptm}\selectfont Non-IID Dataset}} & \multirow{2}{*}{\textbf{\fontfamily{ptm}\selectfont Privacy Leakage}} \\
		\cmidrule(lr){2-5} \cmidrule(lr){6-9}
		& Accuracy & Precision & Recall & F1-score & Accuracy & Precision & Recall & F1-score & \\
		\midrule
		FL Coalition 1-ICBAC & 0.98 & 0.95 & 0.93 & 0.94 & 0.97 & 0.89 & 0.75 & 0.82 & 0.00 \\
		FL Coalition 2-ICBAC & 0.98 & 0.95 & 0.94 & 0.95 & 0.96 & 0.92 & 0.57 & 0.70 & 0.00 \\
		FL Coalition 3-ICBAC & 0.99 & 0.98 & 0.96 & 0.97 & 0.97 & 1.00 & 0.65 & 0.78 & 0.00 \\
		Central Coalition 1 & 0.99 & 0.95 & 0.95 & 0.95 & 0.99 & 0.95 & 0.96 & 0.95 & 0.32 \\
		Central Coalition 2 & 0.99 & 0.97 & 0.96 & 0.96 & 0.99 & 0.96 & 0.96 & 0.96 & 0.53 \\
		Central Coalition 3 & 0.99 & 0.96 & 0.96 & 0.96 & 0.99 & 0.99 & 0.96 & 0.97 & 0.15 \\
		\bottomrule
	\end{tabular}
	}
\end{table*}

Table~\ref{tab:performance_metrics} compares FL models against centralized models trained on aggregated coalition data. Under IID conditions, FL achieves near-parity with centralized training. Under non-IID conditions, centralized models outperform FL, reflecting known limitations of FedAvg under heterogeneous distributions; however, FL maintains strong privacy properties. Centralized models exhibit measurable privacy leakage (0.15--0.53), while FL coalitions achieve zero privacy leakage.

\subsection{Framework Comparison}

Table~\ref{tab:model_comparison} compares ICBAC against recent blockchain-based access control frameworks. ICBAC uniquely supports dynamic promotion and demotion, AI-driven anomaly detection, and privacy-preserving federated learning. These capabilities address critical limitations of existing static and centralized designs, positioning ICBAC as a comprehensive access control solution for decentralized supply chain ecosystems.

\begin{table*}[t]
	\centering
	\fontfamily{ptm}\selectfont 
	\caption{Comparison between ICBAC and recent frameworks.}
	\label{tab:model_comparison}
	\resizebox{1.1\textwidth}{!}{
	\begin{tabular}{l ccccccc}
		\toprule
		& Transparency & Decentralization & Fine-Grained AC & Promotion & Demotion & Intelligence & Privacy Preserving AI \\
		\midrule
		MedTrace \cite{Li2023} & \checkmark & \checkmark & \checkmark & $\times$ & $\times$ & $\times$ & $\times$ \\
		AccessChain \cite{Sarfaraz2023} & \checkmark & \checkmark & \checkmark & $\times$ & $\times$ & $\times$ & $\times$ \\
		ProChain \cite{Li2024} & \checkmark & \checkmark & \checkmark & $\times$ & $\times$ & $\times$ & $\times$  \\
		\cite{Liu2022} & \checkmark & \checkmark & \checkmark & \checkmark & \checkmark & $\times$ & $\times$ \\
		\cite{Hathaliya2024} & \checkmark & \checkmark & \checkmark & $\times$ & $\times$ & $\times$ & $\times$ \\
		\cite{Sohani2024} & \checkmark & \checkmark & $\times$ & \checkmark & $\times$ & $\times$ & $\times$ \\
		ICBAC & \checkmark & \checkmark & \checkmark & \checkmark & \checkmark & \checkmark & \checkmark \\
		\bottomrule
	\end{tabular}
	}
\end{table*}

\section{Conclusion and Future Work} \label{sec:con}

The ICBAC presents a novel, fine-grained approach to secure and privacy-preserving access control in SCM. It integrates Hyperledger Fabric with FL, using private channels and specialized smart contracts to enable decentralized access control. Anomaly detection is powered by AI and enhanced through FL, ensuring zero privacy leakage across coalitions. Additionally, a game theory-based coalition formation mechanism aligns federated clients based on supply chains preferences, improving learning efficiency. In this framework, the FL approach facilitates privacy-preserving model refinement, enabling AI agents to train collaboratively without sharing raw data. The client-selection mechanism is designed to form optimal coalitions that address cross-channel behavioral variations, ensuring core stability and strategy-proofness while using preference-based selection to maintain participant privacy by avoiding the disclosure of sensitive criteria.

Experimental results demonstrate that ICBAC performs competitively compared to recent static blockchain-based frameworks, while also excelling in dynamic features such as user promotion/demotion, intelligence, and privacy-aware AI. Nonetheless, the model shows limitations when applied to non-IID federated learning scenarios and exhibits slightly reduced performance in blockchain read operations. 
Future work will focus on addressing these challenges, especially designing novel aggregation mechanisms or adopting techniques better suited for heterogeneous data distributions in federated settings. 

Additionally, while our current model uses Hyperledger Fabric's multi-channel architecture to demonstrate cross-supply-chain collaboration in consortium settings, an important future direction is extending ICBAC to fully independent cross-blockchain scenarios. This would involve integrating cross-chain interoperability protocols to enable federated learning across supply chains operating on entirely separate blockchain networks. It is a valuable extension for real-world deployments where supply chains may not share common infrastructure.

\section*{Data Availability}
Data and source code are available from the corresponding author upon request.

\section*{Acknowledgements}
During the preparation of this work, the author(s) used Grammarly and ChatGPT in order to enhance the readability and clarity of the article through spelling correction and linguistic improvement. After using this tools, the author(s) reviewed and edited the content as needed and take(s) full responsibility for the content of the publication.

\bibliographystyle{unsrt}  
\bibliography{references}  


\end{document}